\documentclass[prd, 10pt,aps,nofootinbib, floatfix, notitlepage,twocolumn,superscriptaddress]{revtex4-1}

\usepackage{amsmath,amsfonts,amssymb}
\usepackage{mathtools}
\usepackage{mathrsfs}
\usepackage{graphicx}
\usepackage[english]{babel} 
\usepackage{hyperref} 
\usepackage[dvipsnames]{xcolor}
\usepackage{adjustbox}
\usepackage[normalem]{ulem}
\usepackage[capitalize]{cleveref}
\usepackage{csquotes}
\usepackage{xspace}
\usepackage{slashed}
\usepackage{bigints}

\newcommand{\ee}{\end{equation}}
\newcommand{\bes}{\begin{equation*}}
\newcommand{\ees}{\end{equation*}}

\newcommand{\relpow}[3]{\ensuremath{\left(\frac{#1}{#2}\right)^{#3}}}

  \usepackage[utf8]{inputenc}

\newcommand{\LCDM}{$\Lambda$CDM}
\newcommand{\ACT}{ACT DR4}
\newcommand{\SHOES}{S$H_0$ES}
\newcommand{\final}[1]{#1}

\begin{document}

\title{The weak, the strong and the ugly -- \\ A comparative analysis of interacting stepped dark radiation}

\author{Nils Sch\"oneberg}
\email{nils.science@gmail.com}
\affiliation{Institut de Ciències del Cosmos (ICCUB), Facultat de F\'isica, Universitat de Barcelona (IEEC-UB), Mart\'i i Franqués, 1, E08028 Barcelona, Spain}
\author{Guillermo Franco Abell\'an}
\affiliation{GRAPPA Institute, Institute for Theoretical Physics Amsterdam, University of Amsterdam, Science Park 904, 1098 XH Amsterdam, The Netherlands}
%\email{guillermo.franco-abellan@umontpellier.fr}
\author{Th\'eo Simon}
\affiliation{Laboratoire Univers \& Particules de Montpellier (LUPM), CNRS \& Universit\'e de Montpellier (UMR-5299), Place Eug\`ene Bataillon, F-34095 Montpellier Cedex 05, France}
%\email{theo.simon@umontpellier.fr}
\author{Alexa Bartlett}
\affiliation{Department of Physics and Astronomy, Swarthmore College, Swarthmore, PA 19081, USA}
%\email{abartle1@swarthmore.edu}
\author{Yashvi Patel}
\affiliation{Department of Physics and Astronomy, Swarthmore College, Swarthmore, PA 19081, USA}
%\email{ypatel1@swarthmore.edu}
\author{Tristan L.~Smith}
\affiliation{Department of Physics and Astronomy, Swarthmore College, Swarthmore, PA 19081, USA}
%\email{tsmith2@swarthmore.edu}

\date{\today}

\begin{abstract}

Models which address both the Hubble and $S_8$ tensions with the same mechanism generically cause a pre-recombination suppression of the small scale matter power spectrum. Here we focus on two such models. Both models introduce a self-interacting dark radiation fluid scattering with dark matter, which has a step in its abundance around some transition redshift. In one model, the interaction is weak and with all of the dark matter whereas in the other it is strong but with only a fraction of the dark matter. The weakly interacting case is able to address both tensions simultaneously and provide a good fit to a the {\it Planck} measurements of the cosmic microwave background (CMB), the Pantheon Type Ia supernovae, and a combination of low and high redshift baryon acoustic oscillation data, whereas the strongly interacting model cannot significantly ease both tensions simultaneously. The addition of high-resolution CMB measurements (ACT DR4 and SPT-3G) slightly limits both model's ability to address the Hubble tension. The use of the effective field theory of large-scale structures analysis of BOSS DR12 LRG and eBOSS DR16 QSO data additionally limits their ability to address the $S_8$ tension. We explore how these models respond to these data sets in detail in order to draw general conclusions about what is required for a mechanism to address both tensions. We find that in order to fit the CMB data the time dependence of the suppression of the matter power spectrum plays a central role.

\end{abstract}

\maketitle

\enlargethispage*{4\baselineskip}
\section{Introduction}
The rising tension that exists between the Hubble parameter $H_0$ as measured locally through the distance ladder and as inferred from observations of the CMB anisotropies assuming the standard cosmological model (consisting of standard baryons, photons, neutrinos, a cosmological constant, and cold dark matter-- i.e.~`$\Lambda$CDM') presents a pressing puzzle for cosmologists. Solutions to this tension that preserve the excellent agreement with late-time observables typically restrict modifications of the cosmological history to times before recombination. These early-time solutions often suffer from a variety of more or less quantifiable issues, such as strong fine-tuning of the underlying parameters, a lack of a complete underlying particle physics model, strong impacts on the light element abundances generated at big bang nucleosynthesis, or a lack of falsifiability with current experiments. Even more concretely, many of these early-time solutions exacerbate the well-known tension between measurements of the amplitude of clustering in weak lensing surveys and its value inferred by the CMB (see e.g. \citep{Schoneberg:2021qvd}) as expressed through the mismatch in the parameter $S_8 = \sigma_8 \sqrt{\Omega_m/0.3}$\,, where $\Omega_m$ is the current total matter abundance, and $\sigma_8$ is  the root mean square of matter fluctuations on a 8 $h^{-1}$Mpc scale. In principle, it would be possible to `rescue' many of these solutions by simply adding a completely unrelated ingredient of late-time interactions. Yet, such additions are typically even harder to motivate in a consistent particle~physics~model. 

A promising model avoiding many of these problems of early universe solutions to the Hubble tension was recently proposed in \citep{Aloni:2021eaq} and is based on a well-motivated simple Wess-Zumino supersymmetric Lagrangian. Moreover, \citep{Joseph:2022jsf} has demonstrated that this early-universe model is able to ease the Hubble and $S_8$ tensions simultaneously without unrelated ingredients, presenting a first hope of easing both tensions simultaneously in a single well-motivated model. Almost simultaneously, \citep{Buen-Abad:2022kgf} has proposed another interaction for the stepped dark radiation model in the context of partially acoustic dark matter, the Stepped Partially Acoustic Dark Matter model.

The main difference between the two proposals in terms of their interactions between the stepped sector and the dark matter is that the Lagrangian proposed in \citep{Joseph:2022jsf} generates a weak interaction that affects all of the dark matter, while the proposal in \citep{Buen-Abad:2022kgf} generates a strong interaction that only affects a small fraction of the dark matter.\footnote{\final{The terms \enquote{weak} and \enquote{strong} here do not refer to the strong and weak nuclear forces, but instead to the timescale of the interaction relative to the Hubble rate. A \enquote{strongly} interacting model here would be one with an interaction timescale faster than the Hubble rate, while a \enquote{weakly} interacting model would be one with an interaction timescale longer than the Hubble rate (see \cref{fig:Gamma_o_H}).}}

The aim of this paper is to investigate whether the interacting stepped dark radiation models can successfully ease both the Hubble and the $S_8$ tensions, and to explore the various physical mechanisms at play. For this purpose, we subject these proposed models of interacting stepped dark radiation to the newest CMB and large scale structure data (in terms of BAO as well as full modeling of the smaller scales).

In \cref{sec:model}, we give a more detailed description of the two models under investigation. In \cref{sec:method}, we describe the analysis method and the data sets that these models will be subjected to, and the derived constraints are presented in \cref{sec:baseline,sec:additional}. Additional variations on the presented models can be found in \cref{sec:alt_priors}. We finally conclude in \cref{sec:conclusions}.
 
\section{Underlying particle physics}\label{sec:model}
\begin{figure}
    \centering
    \includegraphics[width=0.45\textwidth]{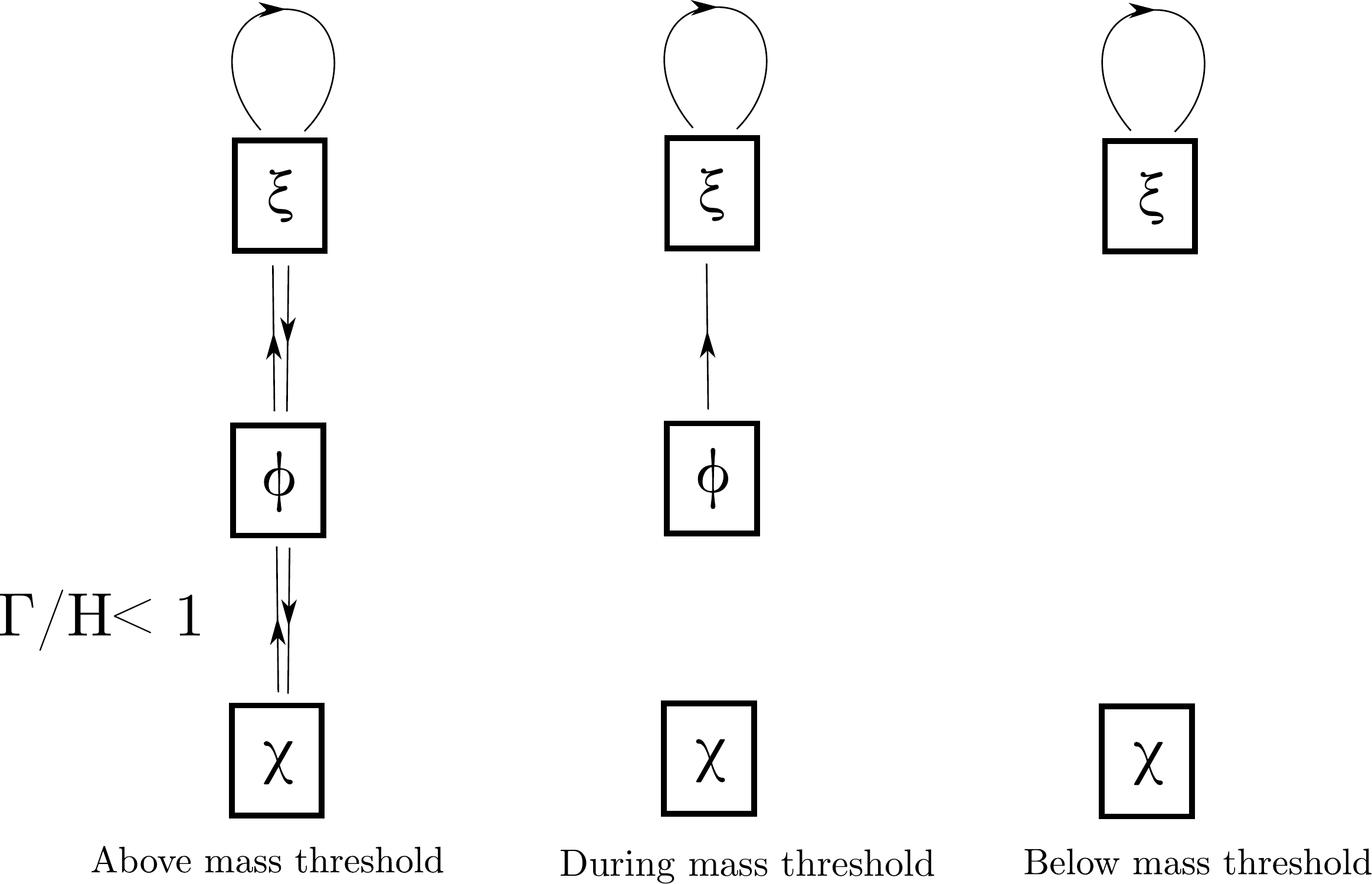}\\
    \rule{0.45\textwidth}{1pt}\\[1em]
    \includegraphics[width=0.45\textwidth]{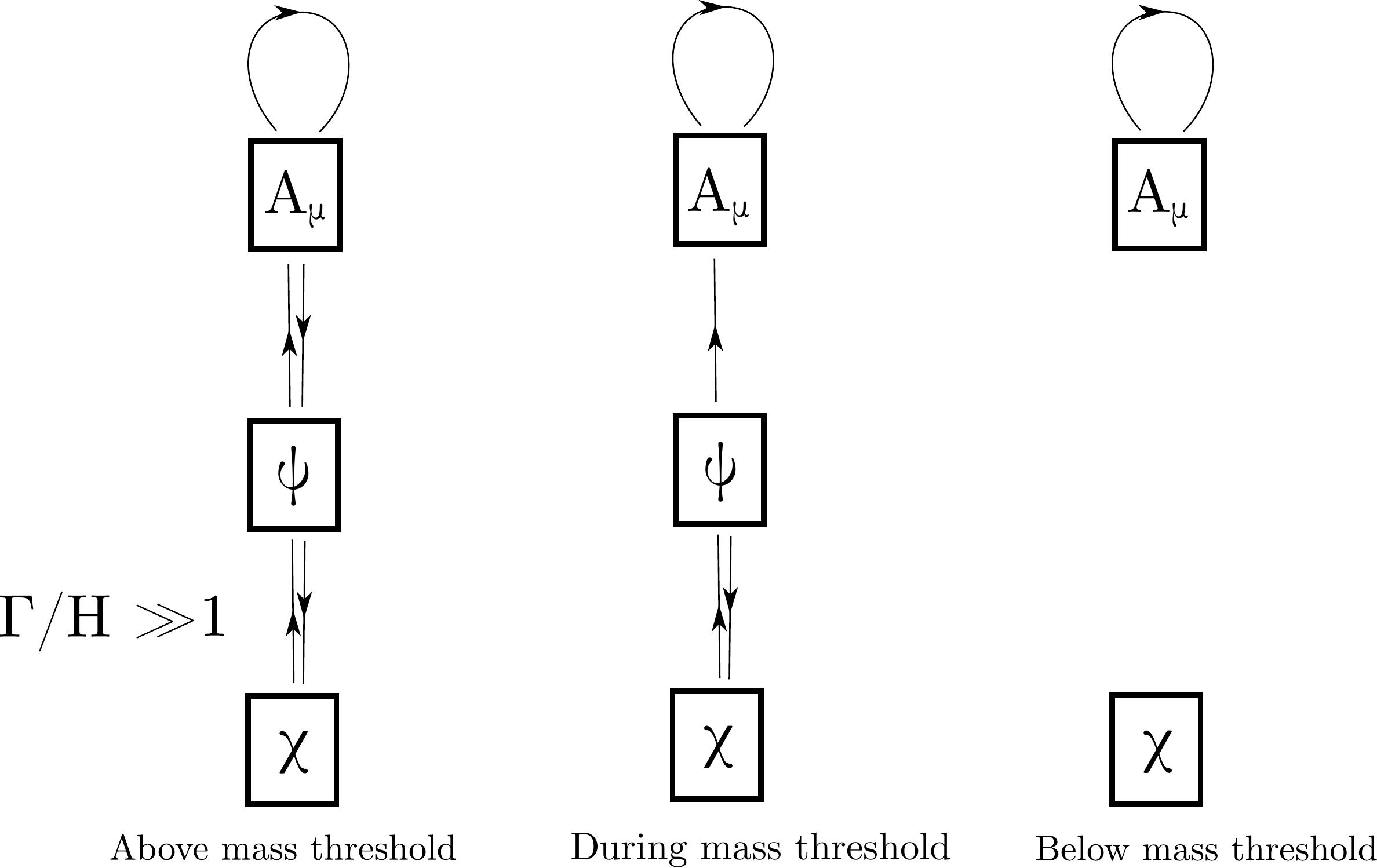}
    \caption{\final{Overview over the two models. Top: Weakly interacting model. Here $\chi$ represents the dark matter, $\phi$ the massive boson (complex scalar), and $\xi$ the massless (Weyl) fermion. Bottom: Strongly interacting model. Here $\chi$ represents the dark matter, $\psi$ the massive fermion, and $A_\mu$ the massless (vector) boson.}}
    \label{fig:overview}
\end{figure}

In this Section, we quickly summarize the most important details of the interacting stepped dark radiation model. In \cref{ssec:model_noninteracting} we will introduce the general description of a stepped dark radiation model without interactions, and in \cref{ssec:model_weak} and \cref{ssec:model_strong} we will shortly introduce the weakly and strongly interacting models, respectively, and summarize the main points in \cref{ssec:model_summary}. Finally, in \cref{ssec:impact} we briefly show the cosmological impact of these models, leaving a more detailed discussion of the underlying physics to \cref{app:impact}. Readers well versed in the physics and notation should feel free to skip to \cref{sec:method}. \final{For a schematic overview, see also \cref{fig:overview}.}

\subsection{Non-interacting stepped dark radiation}\label{ssec:model_noninteracting}
The stepped dark radiation model consists of two strongly self-interacting species -- a massless species and a massive species, which becomes non-relativistic around some transition redshift $1+z_t = m/T_{\mathrm{DR},0}$\,, where $m$ corresponds to the mass of the massive species and $T_{\mathrm{DR},0}$ to the temperature of the dark radiation today. While before the transition redshift the annihilation and decays of both particles balance, after the transition redshift the generation of the massive particles is naturally suppressed through the mass gap of the interactions. This both leads to an exponential decline of the massive particle abundance and an increase in abundance and temperature of the massless particle, leading to an overall increase in the effective number of neutrinos $N_\mathrm{eff}(z)$ after the transition redshift (see Fig. 1 of \cite{Aloni:2021eaq}).

This step-like transition allows the model to boost the effects of $N_\mathrm{eff}$ on the smaller CMB multipoles (below around $\ell \simeq 1000$, as these modes enter the Hubble horizon after $z_t$) without severely impacting the high-$\ell$ diffusion damping tail, which would otherwise critically constrain such dark radiation models.
Another reason why the stepped dark radiation models typically are less constrained from CMB observations than free-streaming dark radiation models is due to the strong self-interaction of the dark radiation components, which prevents the dark radiation from strongly suppressing the growth of potential wells on small scales that is commonly seen in free-streaming dark radiation (resulting in \enquote{neutrino drag}). For a summary of the phenomenology of the stepped dark radiation model see \cite{Aloni:2021eaq}.

We mention here explicitly that in this model the dark radiation species are assumed to be created only after big bang nucleosynthesis (BBN), as otherwise very tight constraints on $N_\mathrm{wzdr}$ apply to this model coming from measurements of the light element abundances (see for example \cite{Schoneberg:2022grr}).

Quite naturally the Lagrangian can also be extended by an interaction between the stepped dark radiation components and (a fraction of) the cold dark matter, and we will discuss the specific extensions below.

\subsection{Weakly interacting stepped dark radiation} \label{ssec:model_weak}
In the weakly interacting stepped dark radiation model introduced in \cite{Aloni:2021eaq,Joseph:2022jsf}, the massive particle is a boson $\phi$, while the massless particle is a fermion. One immediate consequence from entropy conservation is that the increment of the effective neutrino number is given by $N_\mathrm{DR}(z=0)/N_\mathrm{DR}(z\to\infty) = (15/7)^{1/3} \approx 1.29$ in this model. Another consequence is that the natural interactions in this model arise from a Yukawa coupling of the dark matter with the massive boson (such as $\phi \bar{\chi}\chi$). While the massive boson is still present, the dark matter will (weakly) interact with the self-coupled dark radiation fluid, but when the massive boson decays/annihilates away, the interaction of the massless fermion with the dark matter can only occur through the exchange of virtual massive bosons. In this regime, due to the high mass of the virtual particles compared to the thermal energy of the massless fermions, the interaction is suppressed by a factor of $(T_{\mathrm{DR}}/m)^4$ which forces the interaction rate to decline faster than the Hubble rate. 

We also note that in this model one has $\Gamma/H \approx \mathrm{const}$ at early times before the transition redshift, where $H$ is the Hubble parameter. If one then sets the interaction strength such that $\Gamma/H < 1$ at early times, one finds very similar dynamics to non-Abelian dark radiation interacting with dark matter \cite{Buen-Abad:2015ova,Buen-Abad:2017gxg,Lesgourgues:2015wza} (equivalent to the ETHOS $n=0$ case of \cite{Cyr-Racine:2015ihg,Archidiacono:2017slj,Archidiacono:2019wdp,Becker:2020hzj}), which has been shown in the past to have a potentially very strong impact on the clustering of structures. In particular, the suppression of the clustering is an integrated effect, leading to a comparatively smooth/shallow suppression of the power spectrum (as we expand upon in \cref{app:impact_pm}).

\subsection{Strongly interacting stepped dark radiation} \label{ssec:model_strong}
In the strongly interacting stepped dark radiation model introduced in \cite{Buen-Abad:2022kgf}, the massive particle is a fermion instead, while the massless particle is a vector boson $A_\mu$\,. In this case, the increment of the dark radiation effective neutrino number is $N_\mathrm{DR}(z=0)/N_\mathrm{DR}(z\to\infty) = (11/4)^{1/3} \approx 1.4$, giving a slightly larger step than in the weakly interacting case. The dark matter interaction is generated through a coupling from the covariant derivative $D_\mu = \partial_\mu - i g A_\mu$, either with fermionic dark matter $\bar{\chi}(i\slashed{D}-m_\chi)\chi$ or with bosonic dark matter as $|D\chi|^2$. In either case, the massless vector boson mediates the interaction. However, direct scattering (e.g.~Comtpon-like) of the massless species with the dark matter is only possible through virtual exchanges of dark matter particles, which are heavily penalized if the dark matter mass is reasonably large. Instead, the most efficient scattering occurs from the t-channel process of the massive fermion exchanging a virtual boson with the dark matter. Naturally, this scattering is exponentially suppressed as the massive fermion decays away, effectively receiving a penalty factor of $\exp(-m/T_{\mathrm{DR}})$\,. In this case, due to the choice of a vector boson (as opposed to a scalar boson), there is an additional concern of keeping the remaining massless fermions self-interacting \cite{Buen-Abad:2022kgf}. However, we will restrict ourselves to parameter ranges where such strong self-interaction is guaranteed except where explicitly stated otherwise. 

Interestingly, this requirement puts us into a regime where the interaction rate is much stronger than the Hubble rate $\Gamma \gg H$ before the transition redshift. This, in turn, requires us to postulate only a subdominant fraction of the overall dark matter to be interacting (which we denote by $f_{\rm idm}$), as otherwise, the suppression of clustering would be too strong and observable both in the CMB and LSS, as also evident from the constraints in \cref{sec:baseline}. Since the interaction is significantly larger than the Hubble rate before the transition redshift, the time until the exponential term suppresses $\Gamma/H \sim 1$ in this model is typically delayed by around one decade in redshift (see also \cref{fig:Gamma_o_H}).

\subsection{Summary of interacting dark radiation} \label{ssec:model_summary}

To summarize, both the strongly and the weakly interacting stepped dark radiation models are founded on the mechanism of the dark radiation transition introduced by a dark sector containing a massive and massless species. The main difference arising directly from the chosen Lagrangians is how strong the interaction is before the step ($\Gamma/H \ll 1$ in \cite{Aloni:2021eaq,Joseph:2022jsf}, and $\Gamma/H \gg 1$ in \cite{Buen-Abad:2022kgf}) -- thus also justifying our naming convention -- as well as how quickly the interaction decays after the step, either polynomially as $(T_{\mathrm{DR}}/m)^4$ or exponentially with $e^{-m/T_{\mathrm{DR}}}$\,. The weakly interacting case also causes a smoother suppression of the power spectrum and negligible dark acoustic oscillations (see \cref{app:impact} for more details).

\begin{figure}[!ht]
    \centering
    \includegraphics[width=\columnwidth]{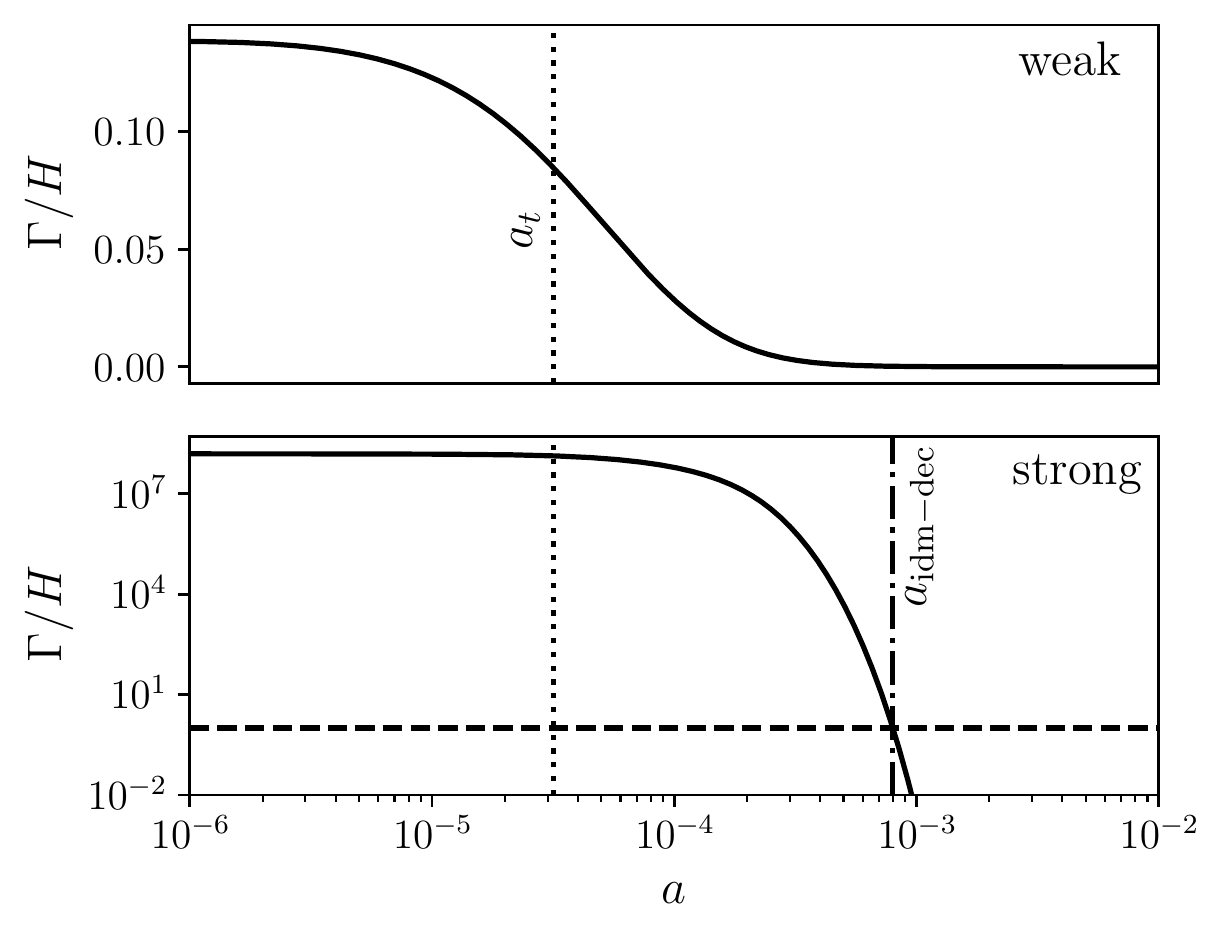}
    \caption{\emph{Top panel}: The interaction rate in units of the Hubble parameter in the weakly interacting case, with $a_t = 10^{-4.5}$ shown in the dotted line. \emph{Bottom panel}: The interaction rate in units of the Hubble parameter in the strongly interacting case, with $a_t = 10^{-4.5}$ shown in the dotted line. We also show the scale factor when the idm decouples ($\Gamma/H = 1$). Note that $a_t$ and $a_{\rm idm-dec}$ are separated by about an order of magnitude.}
    \label{fig:Gamma_o_H}
\end{figure}

We show the behavior of the dimensionless interaction rate $\Gamma/H$ in \cref{fig:Gamma_o_H}. In both models, the interaction rate starts to drop around $a_t$\,. However, in the strongly interacting case (bottom panel) we can see that the decoupling of the dark matter, which occurs when $\Gamma(a_{\rm idm,dec}) = H(a_{\rm idm,dec})$, is delayed by about one order of magnitude in scale factor. 

The general equations for the non-interacting stepped dark radiation model can be found for example in \cite{Aloni:2021eaq,Schoneberg:2022grr}. We modify the Euler equation for the bulk velocity as
\begin{align}
    \dot{\theta}_\mathrm{idm} &= \ldots + a \Gamma (\theta_\mathrm{DR} - \theta_\mathrm{idm})~,\label{eq:euler_idm} \\
    \dot{\theta}_\mathrm{DR} &= \ldots + a \Gamma S (\theta_\mathrm{idm} - \theta_\mathrm{DR})~,\label{eq:euler_wzdr}
\end{align}
where the dot denotes a derivative with respect to conformal time, $S = \rho_\mathrm{idm}/(\rho_\mathrm{DR} + P_\mathrm{DR})$, and 
\begin{equation}
    \Gamma^\mathrm{weak} =\frac{\Gamma_0 ((1+z_t)/x)^2} {(1-0.05 \sqrt{x} + 0.131x)^{4}},
\end{equation}
for the weakly interacting model with $x=m/T_{\mathrm{DR}}$\,, whereas 
\begin{equation}
    \Gamma^\mathrm{strong} =\frac{4\alpha_d^2}{3\pi} \ln(\vartheta) \frac{T_{\mathrm{DR}}^2}{m_\chi} \exp(-x) [2+x(2+x)], \label{eq:strong_Gamma}
\end{equation}
for the strongly interacting model. Here we used the cold dark matter mass\footnote{To be differentiated from the mass of the massive stepped dark radiation particle, $m$.} $m_\chi$\, (fixed to 1000 $\mathrm{GeV}$ for this work) and defined $\vartheta =  K_2(x) (x K_0(x) + K_1(x))^{-2} \cdot (\pi \alpha_d^{-3})/(2 g_\psi)$\, (with $g_\psi$ being the degrees of freedom for the massive fermion). The $K_i(x)$ are the Bessel-K functions and $\Gamma_0$ and $\alpha_d$ are the interaction strength parameters of each model. The derivation of these expressions for $\Gamma$ can be found in \cite[Eq.~(3)]{Joseph:2022jsf} for the weakly interacting model, and in \cite[Eq.~(3.13)]{Buen-Abad:2022kgf} for the strongly interacting model. We will further fix $\alpha_d = 10^{-4}$ in order to ensure a tightly coupled dark radiation species, and instead vary the fraction of dark matter that interacts with the dark radiation, denoted as $f_\mathrm{idm}$\,, as has been done in \cite{Buen-Abad:2022kgf}. Finally, we neglect the impact of interactions on the interacting dark matter sound speed, since the latter impacts primarily small scales beyond the reach of current experiments.

We also implement the initial conditions in a consistent way (detailed in \cref{app:initial}), and take care to stay in the parameter space in which our modeling holds (see \cref{app:limits_SD,app:selfcouple}).\pagebreak[20]

\subsection{Impact on observables}\label{ssec:impact}
In this Section we briefly discuss the impact on cosmological observables of the weakly and the strongly interacting model. However, we leave a more detailed description of the underlying physical mechanisms of the power spectrum suppression and the impact on the CMB to \cref{app:impact}.

\begin{figure}[t]
    \centering
    \includegraphics[width=\columnwidth]{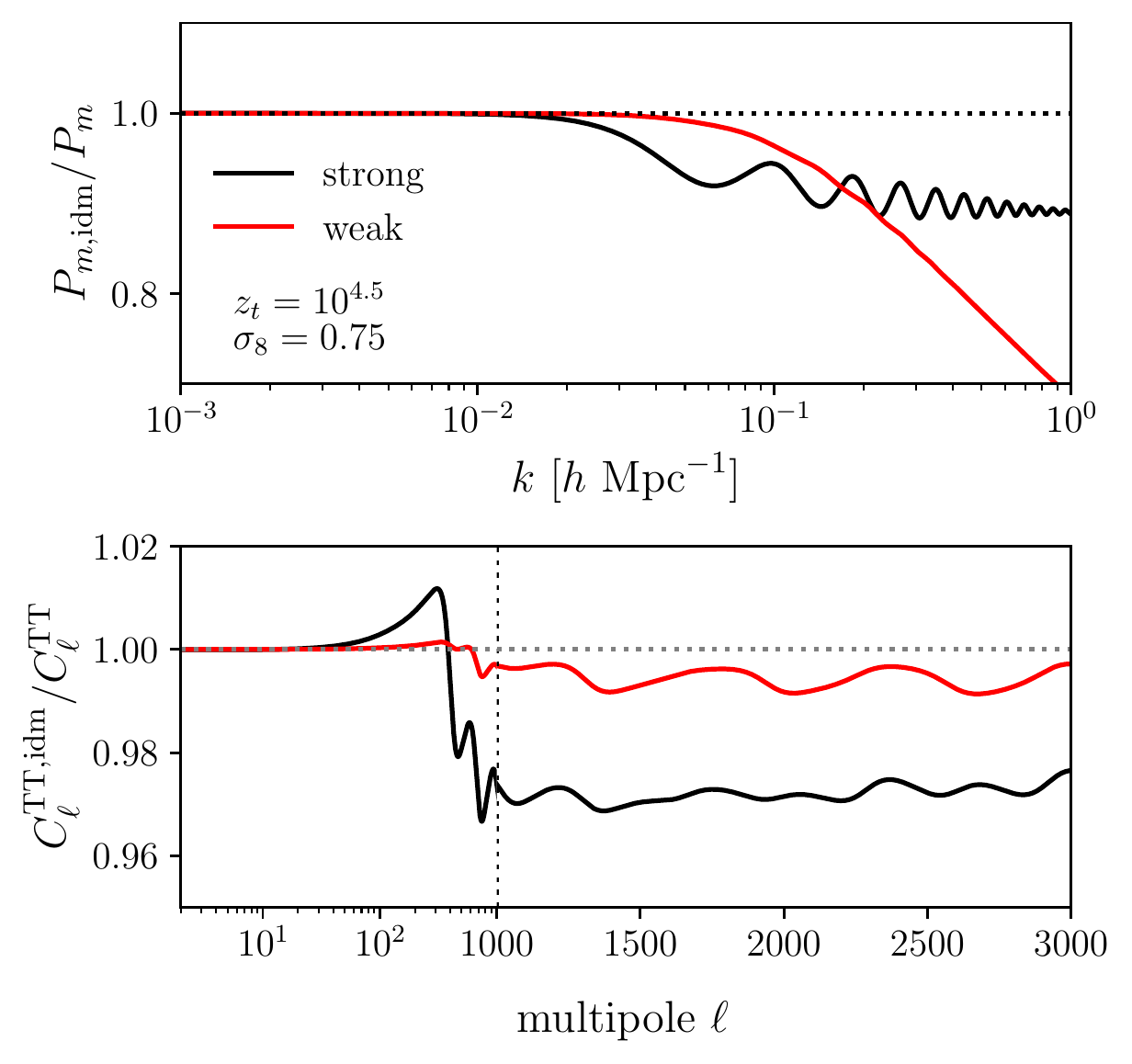}
    \caption{Power spectra (unlensed) for the weakly and strongly interacting models, divided by their non-interacting limits. The parameters of the two model are given by the $\Lambda$CDM bestfit parameters from \cite{Planck:2018vyg} with $N_\mathrm{DR}=0.5, z_t = 10^{4.5}$ and the $f_\mathrm{idm}/\Gamma_0$ adjusted to give $\sigma_8 = 0.75$}
    \label{fig:Pm}
\end{figure}

The impact of each model on the matter power spectrum and the unlensed TT angular power spectrum is shown in \cref{fig:Pm}. The parameters of the model have been chosen such that the impact on the Hubble constant is large ($z_t=10^{4.5}, N_\mathrm{DR}=0.5$) while also simultaneously leading to a small $S_8$ value ($\sigma_8=0.75$).

We observe that the suppression in the weak model is smoother but continues to increase in amplitude for large wavenumbers $k$, whereas the suppression of the strong model has a step-like feature around the modes corresponding to the decoupling redshift, with the suppression only growing slowly for very large $k$ or $\ell$ (not visible in the shown range). In the weakly interacting model the suppression of power is accumulative (as $\Gamma/H \ll 1$) and is proportional to the ratio of the scale factor between Hubble entry and decoupling. Instead, in the strongly interacting model (with $f_{\rm idm} \ll 1$) all modes that enter the Hubble horizon between matter/radiation equality and the interacting dark matter decoupling, are suppressed due to the strong interaction between the dark matter and dark radiation, leading to a drop in power as the fluctuations in the interacting component are essentially erased. This also leads to the strong dark acoustic oscillations in the matter power spectrum shown in the top panel of \cref{fig:Pm}. 

\section{Analysis method and data sets}\label{sec:method}

For the numerical evaluation of the cosmological constraints on the models considered within this work and their statistical comparison we perform a series of Markov-chain Monte Carlo (MCMC) runs using the public code {\sf MontePython-v3}\footnote{\url{https://github.com/brinckmann/montepython_public}} \citep{Audren:2012wb,Brinckmann:2018cvx}, interfaced with our modified versions of {\sf CLASS}\footnote{\url{https://lesgourg.github.io/class_public/class.html}} \cite{Lesgourgues:2011re,Blas:2011rf}. We make use of a Metropolis-Hasting algorithm assuming flat priors on $\{\omega_b,\omega_{\rm cdm},H_0,\ln(10^{10}A_s),n_s,\tau_{\rm reio}\} $\footnote{Here $\omega_b$ and $\omega_{\rm cdm}$ are the physical baryon and cold dark matter energy densities, respectively, $A_s$ is the amplitude of the scalar perturbations, $n_s$ is the scalar spectral index, and $\tau_{\rm reio}$ is the reionization optical depth.}. When considering the two interacting stepped radiation models we also vary the amount of tightly coupled dark radiation, $N_{\rm DR}$, the logarithm of the redshift at which the step occurs, $\log_{10}(z_t)$, and either the dark matter-dark radiation interaction rate ($\Gamma_0$) for the weakly interacting model or the fraction of the dark matter which is tightly coupled to the dark radiation ($f_\mathrm{idm}$) for the strongly interacting model. We use flat priors on these parameters, constraining $N_\mathrm{DR}>0$ as well as\footnote{The prior of $\log_{10}(z_t)$ is purposefully chosen larger than the [4,4.6] range adopted in \cite{Aloni:2021eaq,Joseph:2022jsf}, since the strongly interacting model has interesting features outside of this range and we aim to put both models on the same footing. Additionally, strongly constraining the model can artificially aid its power in decreasing the Hubble tension at the cost of a certain level of fine-tuning.} $\log_{10}(z_t) \in [3,5]$ in order to remain within the cosmologically relevant region.

We adopt the {\em Planck} collaboration convention in modeling free-streaming neutrinos as two massless species and one massive with $m_\nu=0.06$ eV \cite{Ade:2018sbj}. We use {\sf Halofit} to estimate the non-linear matter clustering \cite{Smith:2002dz} solely for the purpose of the CMB lensing, and discuss this choice further in  \cref{app:nonlinear}.
We consider chains to be converged using the Gelman-Rubin \citep{Gelman:1992zz} criterion $|R -1|\lesssim0.05$.\footnote{This condition is chosen because of the non-Gaussian (and sometimes multi-modal) shape of the posteriors of the parameters. For all \LCDM{} runs we have $|R -1|<0.01$.} 
To analyze the chains and produce our figures we use {\sf GetDist} \cite{Lewis:2019xzd}, and we obtain the minimal $\chi^2$ values using the same method as employed in \cite{Schoneberg:2021qvd}.

We make use of a variety of likelihoods, detailed in the bullet points below.
\begin{itemize}
    \item \textbf{Planck:} The low-$\ell$ CMB temperature and polarization auto-correlations (TT, EE), and the high-$\ell$ TT, TE, EE data~\cite{Planck:2019nip}, as well as the gravitational lensing potential reconstruction from {\it Planck}~2018~\cite{Planck:2018lbu}.
    \item \textbf{SPT:} The SPT-3G likelihood \cite{SPT-3G:2021eoc}, which has been adapted from the official {\sf clik} format\footnote{\url{https://pole.uchicago.edu/public/data/dutcher21} (v3.0)}.
    \item \textbf{ACT:} The \ACT\  \cite{ACT:2020frw} likelihoods \footnote{\url{https://github.com/ACTCollaboration/pyactlike}}. In analyses that include the full \textit{Planck} TT power spectrum, we removed any overlap with \ACT{} TT up until $\ell = 1800$ to avoid introducing correlations between the two data sets \cite{Aiola:2020azj}.
    \item \textbf{BAO:} We consider low-$z$ BAO data gathered from 6dFGS at $z = 0.106$ \cite{Beutler:2011hx}, SDSS DR7 at $z = 0.15$ \cite{Ross:2014qpa} and BOSS DR12 at ${z = 0.38, 0.51, 0.61}$ \cite{Alam:2016hwk}.
    \item \textbf{EFTofBOSS:} Full-modeling information from BOSS DR12 LRG using the effective field theory of large scale structure (EFTofLSS), cross-correlated with the reconstructed BAO parameters \cite{Gil-Marin:2015nqa}. The SDSS-III BOSS DR12 galaxy sample data and covariances are described in \cite{BOSS:2016wmc,Kitaura:2015uqa}. The measurements, obtained in \cite{Zhang:2021yna}, are from BOSS catalogs DR12 (v5) combined CMASS-LOWZ~\footnote{\url{https://data.sdss.org/sas/dr12/boss/lss/}}~\cite{Reid:2015gra}, and are divided in redshift bins LOWZ, $0.2<z<0.43 \  (z_{\rm eff}=0.32)$, and CMASS, $0.43<z<0.7  \ (z_{\rm eff}=0.57)$, with north and south galactic skies for each, respectively denoted NGC and SGC. From these data we use the monopole and quadrupole moments of the galaxy power spectrum. The theory prediction and likelihood for the full-modeling information are made available through {\sf PyBird} \cite{DAmico:2020kxu}.
    \item \textbf{EFTofeBOSS:} The EFTofLSS analysis \cite{Simon:2022csv} of eBOSS DR16 QSOs \cite{eBOSS:2020yzd}. The QSO catalogs are described in \cite{Ross:2020lqz} and the covariances are built from the EZ-mocks described in \cite{Chuang:2014vfa}. There are about 343 708 quasars selected in the redshif range $0.8<z<2.2$, with $z_{\rm eff}=1.52$, divided into two skies, NGC and SGC~\cite{Beutler:2021eqq,Hou:2020rse}. From these data we use the monopole and quadrupole moments of the galaxy power spectrum. The theory prediction and likelihood for the full-modeling information are made available through {\sf PyBird}.
    \item \textbf{Pantheon:} The Pantheon catalog of uncalibrated luminosity distance of type Ia supernovae (SNeIa) in the range ${0.01<z<2.3}$~\cite{Scolnic:2017caz}. We have checked that using the newer Pantheon+ data from \cite{Brout:2022vxf} does not significantly impact our results. 
    \item  $\boldsymbol{{\mathcal{S}}}$: In some of our analyses we also include priors on $S_8 \equiv \sigma_8 \sqrt{\Omega_m/0.3}$ measured by the 3$\times$2pt weak lensing and galaxy clustering analyses of KiDS-1000$\times$dFLensS+BOSS, $S_8 = 0.766^{+0.020}_{-0.014}$ \cite{Heymans:2020gsg}, and DES-Y3 , $S_8 = 0.775^{0.026}_{-0.024}$ \cite{DES:2021wwk}. To show these $S_8$ constraints in our triangle plots and to compute the Gaussian tension to the combined measurement, we use the simple weighted mean and uncertainty of $S_8^\mathrm{GT} = 0.769^{+0.016}_{-0.012}$\,. 
    \item $\boldsymbol{{\mathcal{H}}}$: At times we also use a Gaussian prior from the late-time measurement of the absolute calibration of the SNeIa from \SHOES\, $M_b = -19.253 \pm 0.027$ \cite{Riess:2021jrx}, corresponding to $H_0 = (73.04\pm1.04)$ km/s/Mpc. We call this prior the `$H_0$ prior' (despite being on $M_b$) since for such early universe models as considered in this work the two viewpoints are essentially equivalent\footnote{The $M_b$ prior is still preferable, since it not only more correctly accounts for possible mild correlations with $\Omega_m$, but it also better preserves the log-Gaussian shape in $H_0$ that has been demonstrated in \cite{Riess:2021jrx}.}.
\end{itemize}
Our baseline combination of data, which we denote as `$\mathcal{D}$', corresponds to the Planck + BAO + Pantheon combination. It is thus equivalent to the baseline data combination employed in \cite{Planck:2018vyg}. The \SHOES\ prior is denoted by `$\mathcal{H}$' and the combination of both $S_8$ priors by `$\mathcal{S}$'. In addition, the  high-resolution \ACT\ and SPT-3G CMB measurements are denoted by `ACT+SPT', while the EFT full-modeling analysis of BOSS and eBOSS data QSO are denoted `EFT'.

We note that the combined analysis of \cite{Kilo-DegreeSurvey:2023gfr} gives an $S_8$ posterior significantly more consistent with the CMB data even in the $\Lambda$CDM model ($S_8 = 0.790^{+0.018}_{-0.014}$, $1.7\sigma$ tension). It remains to be seen if future experiments continue to show an $S_8$ tension. In this work, we will continue using the aforementioned $\mathcal{S}$ priors in order to investigate the respective models' abilities to ease a possible current/future tension.

\section{Baseline data}\label{sec:baseline}

\begin{figure*}
    \centering
    \includegraphics[width=\textwidth]{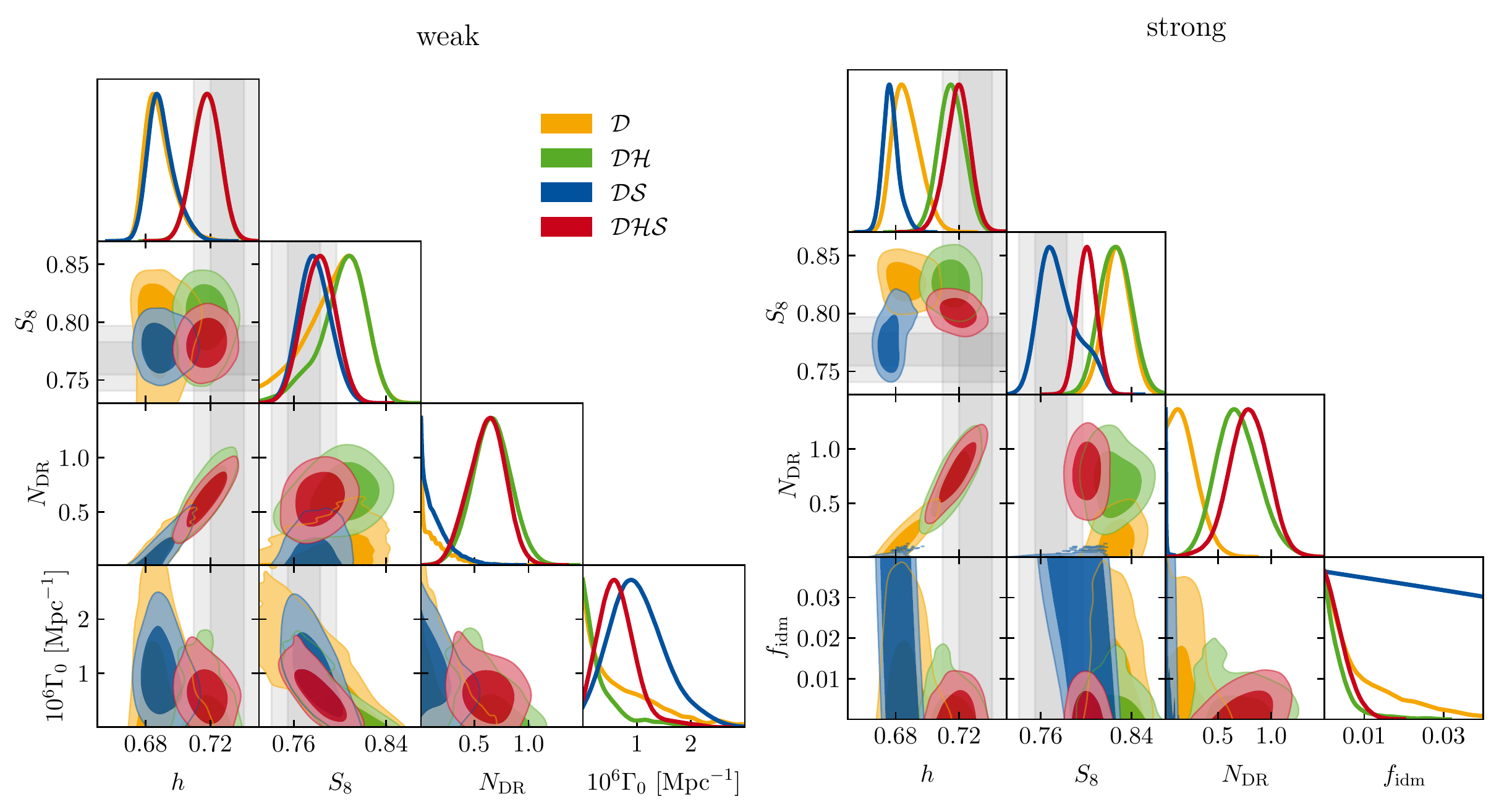}
    \caption{Triangle plots of the two-dimensional constraints (68\% and 95\% C.L.) for the strongly (left) and weakly (right) interacting models for the baseline data set and a variety of added priors (see legend). Particularly notable is the similarity in the $h - N_\mathrm{DR}$ panel (caused by the same underlying stepped dark radiation mechanism), the anti-correlation in the $N_\mathrm{DR}-f_\mathrm{idm}$/ $N_\mathrm{DR}-\Gamma_0$ planes between all combinations, and the important difference between the two models in the $h-S_8$ plane. The gray bands show the $H_0$ and $S_8$ priors.}
    \label{fig:summary_constraints}
\end{figure*}

In \cref{fig:summary_constraints}, we show the constraints using our baseline data set $\mathcal{D}$, as well as the additional $H_0$ and $S_8$ priors. As expected, since both models are extensions of the stepped dark radiation paradigm, they are both equally able to ease the Hubble tension when a prior on $H_0$ is added. While the range of $H_0$ values without the $H_0$ prior is only moderately broader (see \cref{tab:constraints_baseline}), with the addition of the prior the increased compatibility with high values of $H_0$ is revealed: we obtain $H_0 = (71.52\pm0.80)$km/s/Mpc for the strong model, $(71.73\pm0.84)$km/s/Mpc for the weak model. Indeed, with the $H_0$ prior, non-zero values of $N_\mathrm{DR}$ are preferred in agreement with \cite{Aloni:2021eaq}. Similarly, both interactions are able to strongly decrease the value of $S_8$ once a prior is added ($S_8 = 0.786 \pm 0.020$ for the strong model, $0.778 \pm 0.014$ for the weak model).

However, the main difference between the two models is revealed when both priors are imposed simultaneously. While in the weak model, both a low $S_8$ and high $H_0$ can be reached (see \cref{tab:constraints_baseline}), this is not possible in the strongly interacting model. In this model, the region of high $H_0$ does not allow for low $S_8$ (it is only allowed for low $H_0$, see \cref{fig:summary_constraints}), and thus the more constraining $H_0$ prior forces the model to remain at relatively large values of $S_8$ even against the opposing $S_8$ prior. It is interesting to note that due to the weaker constraint on the weak interaction than the strong interaction, even without the $S_8$ prior, the weakly interacting model has a smaller value of $S_8 = 0.802\pm 0.019$ more compatible with DES/KiDS data (compared to $S_8 =  0.818\pm 0.013$ for the strong model).

\begin{table*}
    \centering
    \resizebox{\textwidth}{!}{
    \begin{tabular}{c|c c c c| c c c c}
         & \multicolumn{4}{c | }{strongly interacting model}& \multicolumn{4}{c}{weakly interacting model}\\
        \rule{0pt}{4ex} Parameter & $\mathcal{D}$  & $\mathcal{DH}$ & $\mathcal{DS}$& $\mathcal{DHS}$& $\mathcal{D}$  & $\mathcal{DH}$ & $\mathcal{DS}$ & $\mathcal{DHS}$\\ \hline
       \rule{0pt}{3ex}     $H_0$[km/s/Mpc] & $68.69\pm0.85$ & $71.52 \pm 0.80$ & ${68.10}^{+0.78}_{-0.69}$ & ${71.90}^{+0.79}_{-0.82}$ & ${68.86}^{+0.87}_{-0.84}$ & $71.73 \pm 0.84$ & ${68.98}^{+0.82}_{-0.79}$ & $71.74 \pm 0.84$\\
        $S_8$ & $0.820\pm 0.013$ & $0.818 \pm 0.013$ & $0.786 \pm 0.020$ & $0.8010 \pm 0.0088$ & $0.786 \pm 0.032$& $0.802 \pm 0.019$ & $0.778 \pm 0.014$ & $0.782 \pm 0.014$\\ \hline
        \rule{0pt}{3ex}$10^9A_s$ & $2.112\pm 0.032$ & $2.113 \pm 0.033$ & $2.109\pm0.032$ & $2.091 \pm 0.032$ & $2.096 \pm 0.033$ & $2.115 \pm 0.032$ & $2.090 \pm 0.032$ & $2.114 \pm 0.033$\\
        $n_s$ & $0.9735 \pm 0.0062$ & ${0.9781}^{+0.0089}_{-0.0085}$ & ${0.9707}^{+0.0054}_{-0.0051}$ & $0.9747 \pm 0.0053$ & $0.9729 \pm 0.0050$ & ${0.9819}^{+0.0071}_{-0.0077}$ & $2.090 \pm 0.032$ & $0.9863 \pm 0.0058$\\
        $\Omega_m$ & $0.3099 \pm 0.0063$ & $0.2975\pm 0.0053$ & ${0.3097}^{+0.0068}_{-0.0072}$ & $0.2926 \pm 0.0048$ & $0.3060 \pm 0.0061$& $0.2979 \pm 0.0051$ & $0.3048 \pm 0.0059$ & $0.2974 \pm 0.0056$\\
        $\tau_\mathrm{reio}$ & $0.0563 \pm 0.0072$ & $0.0590 \pm 0.0075$ & $0.0562 \pm 0.0074$ & $0.0571 \pm 0.0075$ & $0.0540 \pm 0.0074$ & ${0.0584}^{+0.0073}_{-0.0075}$ & $0.0528 \pm 0.0072$ & $0.0568 \pm 0.0073$\\ \hline
        \rule{0pt}{3ex} $N_\mathrm{DR}$ & $<0.45$ & $0.68\pm 0.20$ & $< 0.23$ & $0.74 \pm 0.21$ & $< 0.47$ & $0.66 \pm 0.17$ & $< 0.44$ & $0.62 \pm 0.17$\\
        $\Gamma_0$ [$10^{-6}$/Mpc] & -- & -- & -- & -- & $< 3.0$ & $< 1.2$ & $1.04 \pm 0.51$ & $0.65^{+0.31}_{-0.34}$\\
        $f_\mathrm{idm}$ & $<0.039$ & $< 0.015$ & $< 0.30$ & $< 0.0096$& -- & -- & -- & --\\
    \end{tabular}
    }
    \caption{The mean and $1 \sigma$ limits (optionally $95\%$ C.L.~upper limit) for the different parameters for the weak and the strong model, given the baseline data set $\mathcal{D}$ and either the $H_0$ prior ($\mathcal{H}$) or $S_8$ prior ($\mathcal{S}$). See \cref{app:chi2} for a table of corresponding bestfit and minimized chi squared values.}
    \label{tab:constraints_baseline}
\end{table*}

This difference in behavior between these otherwise so similar models solicits a more detailed investigation, which can be found in \cref{app:impact}. The main conclusion is that this different behavior is a direct consequence of the different underlying particle physics models. The weakly interacting model predicts a relatively smooth suppression in $k$ which is in place well-before matter/radiation equality, $a_{\rm eq}$. On the other hand the strongly interacting model causes suppression mainly after $a_{\rm eq}$ leading to additional driving of the photon perturbations which is imprinted on the CMB. Thus, when both $N_\mathrm{DR}$ is large (to ease the $H_0$ tension) a large fraction of interacting dark matter (to ease the $S_8$ tension) is disallowed due to the strong impact on the CMB. Correspondingly, in \cref{fig:summary_constraints}, we observe that while the interaction parameter $f_\mathrm{idm}$ of the strongly interacting model is always constrained from above, the corresponding parameter $\Gamma_0$ of the weakly interacting model shows a preference for nonzero values when the $S_8$ prior is added.

\begin{figure}
    \centering
    \includegraphics[width=0.9\columnwidth]{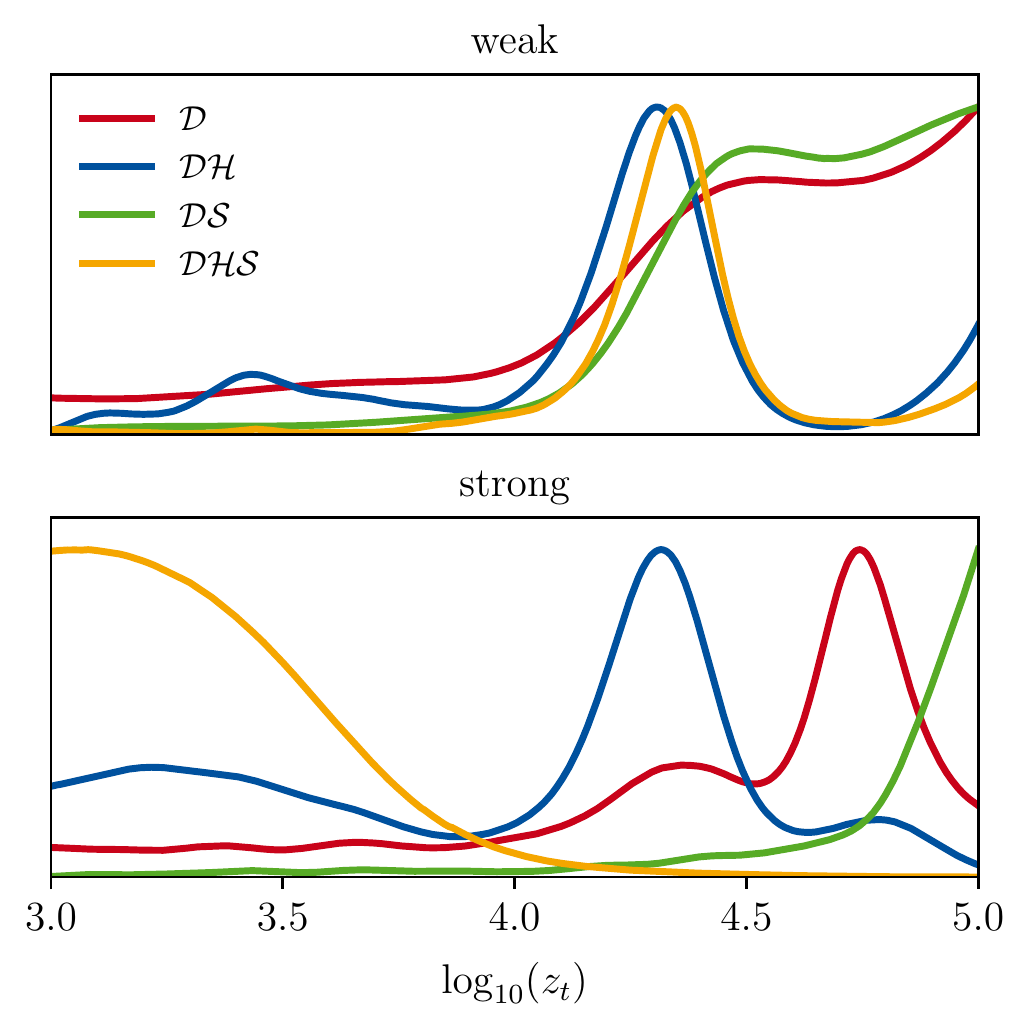}
    \caption{One-dimensional posterior constraints on the parameter $\log_{10}(z_t)$ in the weakly interacting model (top) and the strongly interacting model (bottom) for a variety of data sets (see legend).}
    \label{fig:log10zt}
\end{figure}

This can also be observed in \cref{fig:log10zt}. In order to efficiently ease the Hubble tension, each model should have $\log_{10}(z_t)$ around $4-4.5$ in order to allow for different enhancements of low and high CMB multipoles \cite{Joseph:2022jsf}. The blue curves representing the sole addition of the $H_0$ prior show this behavior for both models. If the $S_8$ prior is added on top (orange curves), the weakly interacting model still has a significant part of the posterior in this regime, while the strongly interacting model prefers small values of $\log_{10}(z_t)$, below around 4 (i.e., after $a_{\rm eq}$). In this range, the strongly interacting model is effectively equivalent to that of self-interacting dark radiation from the perspective of the CMB. We note that \cref{fig:log10zt} differs from Fig.~5 of Ref.~\cite{Aloni:2021eaq} due to the broadening of the prior range.

We show in \cref{tab:tensions_baseline} a comparison of the $\Delta$AIC and Gaussian tension criteria between the models, similarly to the analyses in \cite{Schoneberg:2021qvd,Joseph:2022jsf}. This shows that in all cases the weakly interacting model is indeed able to ease the $S_8$ tension, and mostly performs just slightly better in terms of the $H_0$ tension. We further observe that the $\Delta$AIC of the weakly interacting model is also improved for all combinations, whereas the strongly interacting model mostly benefits from its ability to ease the Hubble tension (the $\Delta$AIC in the case with only $S_8$ prior is positive). It should be stressed, however, that both models perform remarkably well compared to $\Lambda$CDM when faced with the challenge of reconciling each tension separately.

\begin{table}[h]
    \centering
    \resizebox{\columnwidth}{!}{
    \begin{tabular}{c|c c c c| c c c c}
         & \multicolumn{4}{c | }{strongly interacting model}& \multicolumn{4}{c}{weakly interacting model}\\
        \rule{0pt}{4ex} Tension metric & $\mathcal{D}$  & $\mathcal{DH}$ & $\mathcal{DS}$& $\mathcal{DHS}$& $\mathcal{D}$  & $\mathcal{DH}$ & $\mathcal{DS}$ & $\mathcal{DHS}$\\ \hline
       \rule{0pt}{3ex}  
       Gaussian tension $H_0$ & $3.3\sigma$ & $1.2\sigma$ & $3.9\sigma$ & $0.9\sigma$ & $3.1\sigma$ & $1.0\sigma$ & $3.1\sigma$ & $1.0\sigma$ \\ 
       Gaussian tension $S_8$ & $2.5\sigma$ & $2.4\sigma$ & $0.7\sigma$ & $1.7\sigma$ & $0.5\sigma$ & $1.3\sigma$ & $0.4\sigma$ & $0.6\sigma$ \\ \hline \rule{0pt}{3px}
       $Q_\mathrm{DMAP}$ tension $H_0$ & $3.1\sigma$ & -- & $3.5\sigma$ & -- & $2.9\sigma$ & -- & $3.1\sigma$ & -- \\
       $Q_\mathrm{DMAP}$ tension $S_8$ & $1.9\sigma$ & $2.6\sigma$ & -- & -- & $0.9\sigma$ & $1.5\sigma$ & -- & -- \\ \hline \rule{0pt}{3px}
        $\Delta \chi^2$ &  -0.7 & -23.6 & -5.3 & -19.8 & -1.9 & -25.9 & -9.35 & -26.5  \\
        $\Delta$AIC &  5.3 & -17.6 & 0.7 & -13.8 & 4.1 & -19.9 & -3.35 & -20.5 \\
    \end{tabular}
    }
    \caption{Gaussian tensions, $Q_\mathrm{DMAP}$ tensions, and differences in the minimized effective $\chi^2$ (and corresponding $\Delta$AIC) for the various data combinations for the weakly and strongly interacting models. The Gaussian tension is evaluated with respect to the result from \cite{Riess:2021jrx} (our $\mathcal{H}$ prior) and the combination of \cite{Heymans:2020gsg,DES:2021wwk}(our $\mathcal{S}$ prior).}
    \label{tab:tensions_baseline}
\end{table}

\section{Additional data}\label{sec:additional}

\enlargethispage*{2\baselineskip}
Given the success of the two interacting models in terms of the Hubble and $S_8$ tension demonstrated in \cref{sec:baseline}, it is crucial to subject the models to additional data to investigate whether these change the conclusions by favoring one model's suppression mechanism over another. In this Section we confront the interacting stepped dark radiation models with various data sets beyond our baseline combination. In particular, in \cref{ssec:add_cmb} we add CMB data from ACT and SPT, while in \cref{ssec:add_lss} we add full-modeling large scale structure data.

\subsection{Small scale CMB measurements}\label{ssec:add_cmb}

Given the critical role played by the small scale CMB measurements in restricting dark radiation models, it is imperative to subject these models to other measurements of the small scale polarization, such as data from the ACT and SPT collaborations. 

\begin{figure}
     \centering
    \includegraphics[width=\columnwidth]{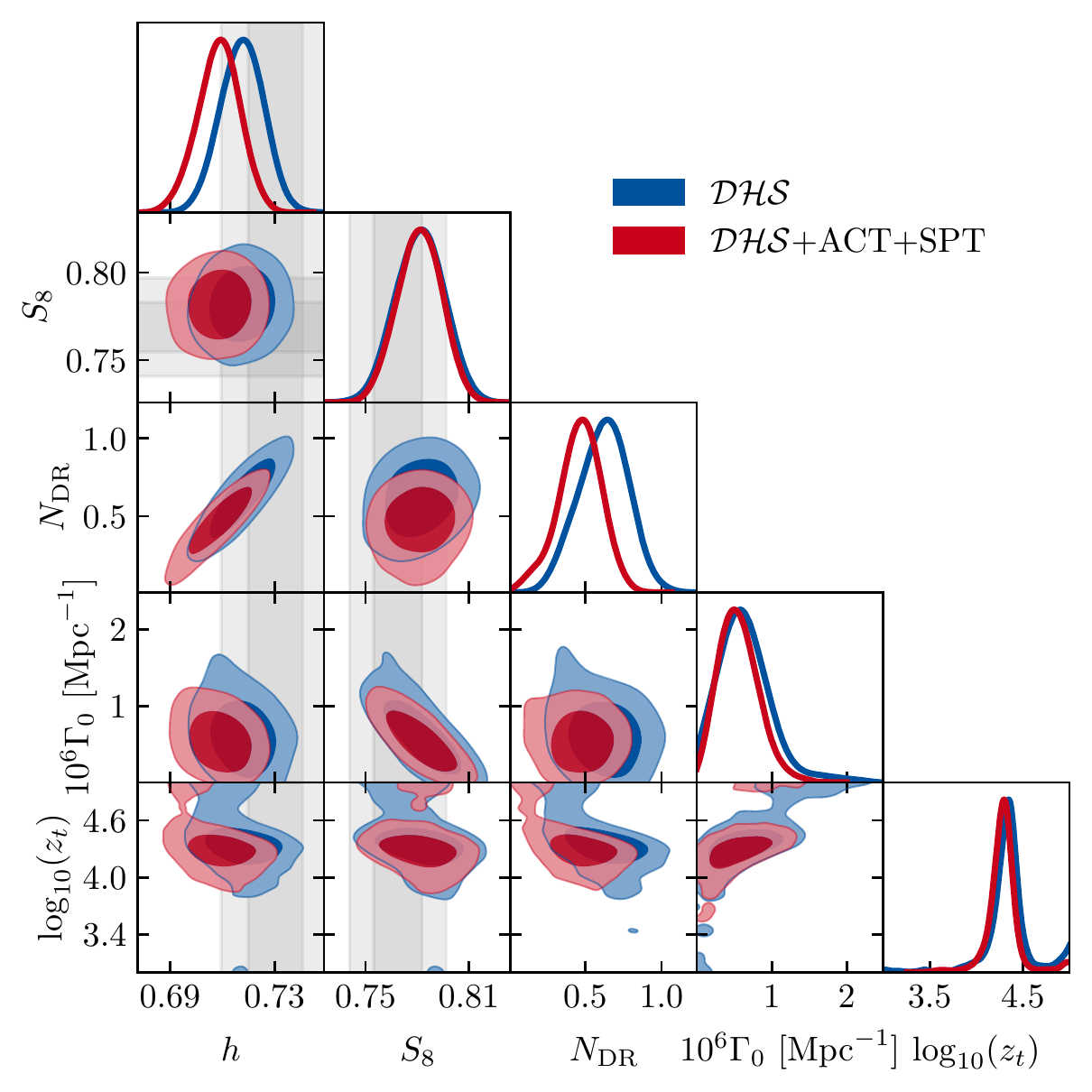}
    \caption{Triangle plot of two-dimensional constraints (68\% and 95\% C.L.) on the parameters of the weakly interacting stepped dark radiation model when confronted with the $\mathcal{DHS}$ data set and the additional ACT+SPT data.}
    \label{fig:ACTSPT}
\end{figure}

We show the results in \cref{fig:ACTSPT} for the weakly interacting model for the data set $\mathcal{DHS}$+ACT+SPT. As expected, we observe a small increase in constraining power and a corresponding shift towards smaller values of $N_\mathrm{DR}=0.46 \pm 0.14$ and correspondingly $H_0=(70.87 \pm 0.81)$km/s/Mpc. Overall, this leads to a slightly increased Hubble tension at the level of $1.6\sigma$ (compare \cref{tab:tensions_baseline}), while the impact on $S_8$ is negligible. Beyond the data from \cite{SPT-3G:2021eoc}, we also checked that the newer SPT data from \cite{SPT-3G:2022hvq} (also including temperature auto-correlation) does not significantly impact the constraints. Indeed, the most notable impact of the newer SPT data appears to be erasing the small upturn of the posterior of $\log_{10}(z_t)$ close to the upper limit visible in \cref{fig:ACTSPT} (and in \cref{fig:log10zt}).

\begin{figure}
    \centering
    \includegraphics[width=\columnwidth]{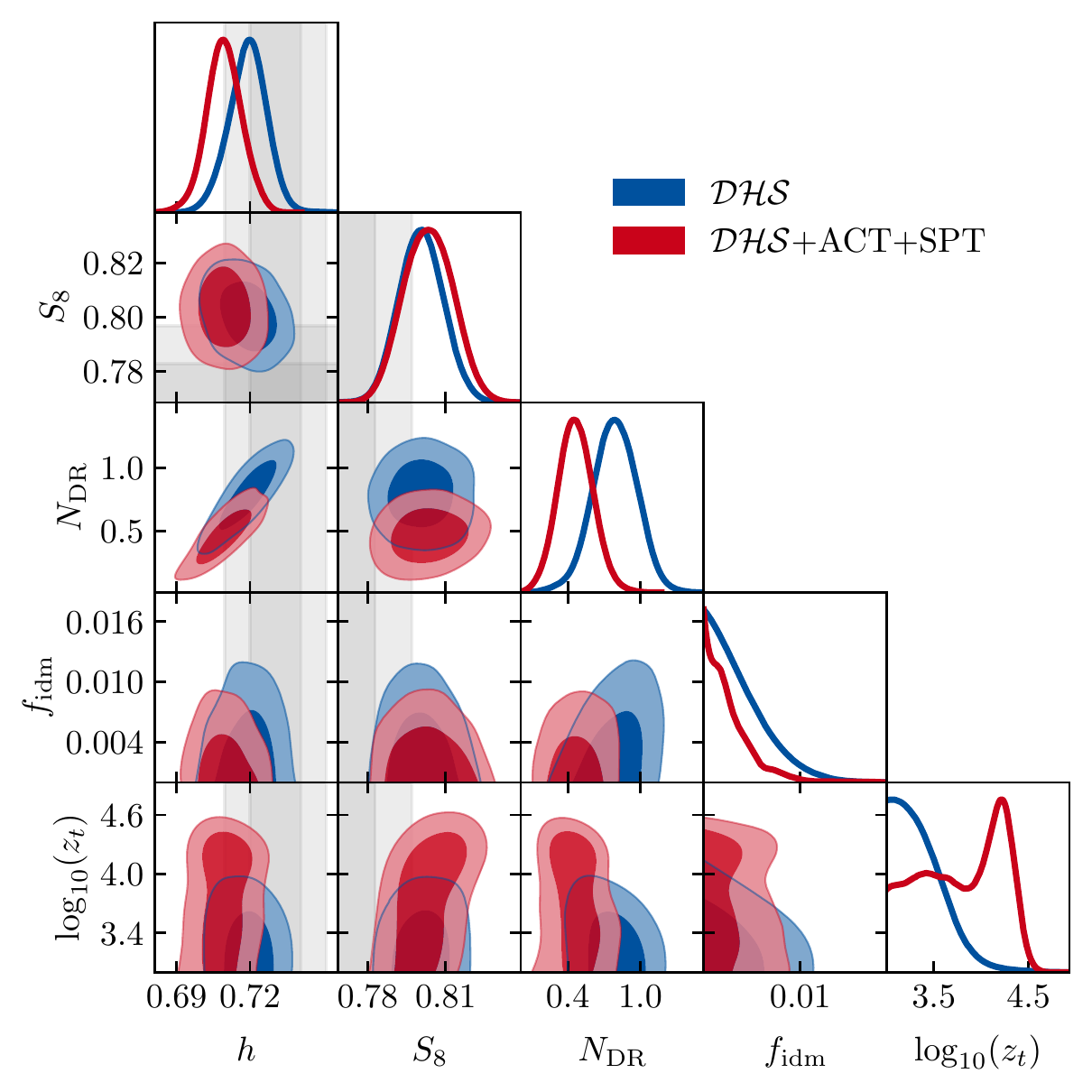}
    \caption{Same as \cref{fig:ACTSPT}, but for the strongly interacting stepped dark radiation model instead.}
    \label{fig:ACTSPT2}
\end{figure}
For the strongly interacting models, the results are shown in \cref{fig:ACTSPT2}. This Figure clearly demonstrates that with these additional data large values of $N_\mathrm{DR}$ are more constrained ($N_\mathrm{DR} = 0.47 \pm 0.14$), with the impact of the ACT+SPT data larger on this model than on the weakly interacting model. Indeed, in this parameter regime of lower $N_\mathrm{DR}$ larger values of $\log_{10}(z_t) \sim 4.5$ are allowed again, while $f_\mathrm{idm}$ is slightly more constrained from the ACT+SPT data ($f_\mathrm{idm}< 0.0078$). The impact on $H_0=(70.95 \pm 0.71)$km/s/Mpc and $S_8=0.8030 \pm 0.0094$ are mild, leading to respectively $1.7\sigma$ and $1.8\sigma$ tension (compare to \cref{tab:tensions_baseline}).

However, while the constraints on the model become tighter and restrict their ability to ease the Hubble tension, the $\Delta {\rm AIC}$ compared to $\Lambda$CDM improves, with the weakly interacting model achieving $\Delta$AIC$=-27.7$ (compared to $-20.5$ without ACT+SPT) and the strongly interacting model achieving $\Delta$AIC$=-18.4$ (compared to $-13.8$ without ACT+SPT). This is caused by each model reducing the tension between ACT+SPT and Planck data, allowing for a better fit to both of them ($\Delta \chi^2_{\rm ACT+SPT+Planck} = -10.8$ for the weakly interacting model, as well as $\Delta \chi^2_{\rm ACT+SPT+Planck} = -10.4$ for the strongly interacting model). The main reason for the worse performance of the strongly interacting model in this metric is its worse performance in terms of BAO data and the $S_8$ prior (see \cref{tab:chi2}). This strong reduction in minimal $\chi^2$ when ACT+SPT datasets are present is particularly interesting since Ref.~\cite{Schoneberg:2022grr} found that these data were rather restrictive to the stepped dark radiation model without interactions, though a formal analysis of the $\Delta$AIC (or other $\chi^2$ metrics) was not performed there. While the significance of this effect is not as large as in Ref.~\cite{Smith:2022hwi} for EDE, we do note the striking similarity. However, a more detailed analysis comparing the ACT, SPT, and Planck residuals in these respective models is left for future work\footnote{We also note that this mechanism is likely not related to the interactions but instead to the underlying stepped dark radiation mechanism. We leave a more detailed analysis for future work.}.

\subsection{Full-modeling galaxy power spectra}\label{ssec:add_lss}

In this Section, we examine the constraints imposed on these two models by the EFT full-modeling analysis of the BOSS+eBOSS galaxy and QSO clustering data (see for example Refs.~\cite{Simon:2022adh, Simon:2022ftd,Moretti:2023drg,DAmico:2020tty,Rubira:2022xhb} for this type of analysis applied to some alternative models). 
Compared to standard BAO+$f\sigma_8$ template-based analyses (adding redshift space distortion information) of BOSS and eBOSS data, the full-modeling constraints from EFT are expected to be much more sensitive to the suppression of the small-scale power spectrum introduced in these models. In addition, in Ref.~\cite{Simon:2022ftd}, it has been shown that even in a model of non-interacting dark radiation the EFT full-modeling analysis alone can put interesting constraints on its abundance. Indeed, we confirmed within the context of our models that the BAO+$f\sigma_8$ data on its own does not impose strong constraints, while for the EFT data alone we find $N_\mathrm{DR} = 3.7_{-3.0}^{+1.2}$ and $f_\mathrm{idm} < 0.10$ for the strongly interacting model, and $N_\mathrm{DR}<7.5$ and $\Gamma_0 < 3.5 \cdot 10^{-6}$/Mpc for the weakly interacting model.

In order to investigate the complimentarity of the EFT and CMB data, we add the EFTofBOSS and EFTofeBOSS likelihoods (see descriptions in \cref{sec:method}) to the data set $\mathcal{D}$, where we have only removed the BOSS DR12 post-reconstructed measurements which are already included in the corresponding {\sf PyBird} likelihood. This new data set is now called `$\mathcal{D}$+EFT'. The $\chi^2$ values for these analyses are provided in \cref{tab:chi2} of \cref{app:chi2}.

Both models provide a good fit to BOSS and eBOSS data at the same time, which means that the EFT parameters are able to compensate for the suppression of the matter power spectrum generated by these two models, which occurs from $k \sim 0.1h/\mathrm{Mpc}$ (see~\cref{fig:Pm}). We note that the EFT data are similarly well fit in these models as in $\Lambda$CDM (see the BOSS and eBOSS $\chi^2$ in \cref{tab:chi2}) and it involves only small ($<1\sigma$) shifts in the nuisance parameters relative to their from their $\Lambda$CDM values. There is a slight worsening in the total $\Delta$AIC with the $\mathcal{DHS}$+EFT data, as the strongly interacting model only achieves $\Delta$AIC=$-8.7$, while the weakly interacting model still achieves $\Delta$AIC=$-19.3$ (compare \cref{tab:tensions_baseline}).

\begin{figure}
    \centering
    \includegraphics[width=0.9\columnwidth]{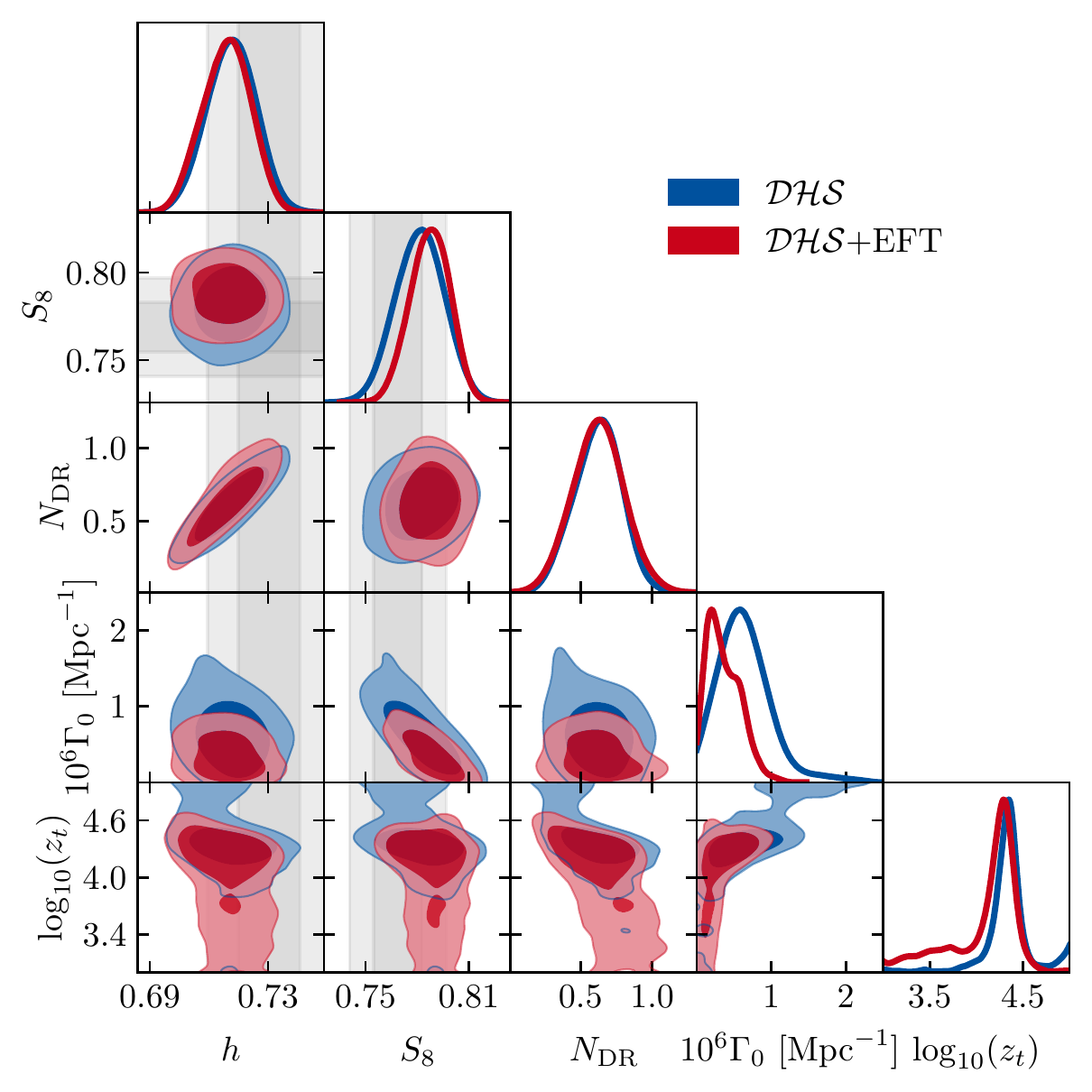}
    \caption{Triangle plot of two-dimensional constraints (68\% and 95\% C.L.) on the parameters of the weakly interacting stepped dark radiation model when confronted with the $\mathcal{DHS}$+EFT data set.}
    \label{fig:eft_weak}
\end{figure}

In terms of parameter constraints for the weak model (shown in \cref{fig:eft_weak}), the main impact of adding EFT analysis to BOSS and eBOSS data is a significantly stronger constraint of the interaction rate by a factor of 3.3 between the $\mathcal{DHS}$ and $\mathcal{DHS}$+EFT analyses (the latter gives $\Gamma_0 = {0.27}^{+0.12}_{-0.08} \cdot 10^{-6}{\rm Mpc}^{-1}$). This new constraint propagates to the $S_8$ parameter ($S_8 = {0.7932}^{+0.0041}_{-0.0066}$), leading to a $+0.7\sigma$ upwards shift and an improvement in constraining power by a factor of 2.6. The associated Gaussian tension of $S_8$ is now $1.4\sigma$ (compared to $0.6\sigma$ without EFT data). The conclusions for the $\mathcal{DS}$ dataset are very similar, with stronger constraints on $\Gamma_0$ leading to higher $S_8$ values, while both $\mathcal{D}$ and $\mathcal{DH}$ combinations are barely affected by the additional EFT data.

\begin{figure}
    \centering
    \includegraphics[width=\columnwidth]{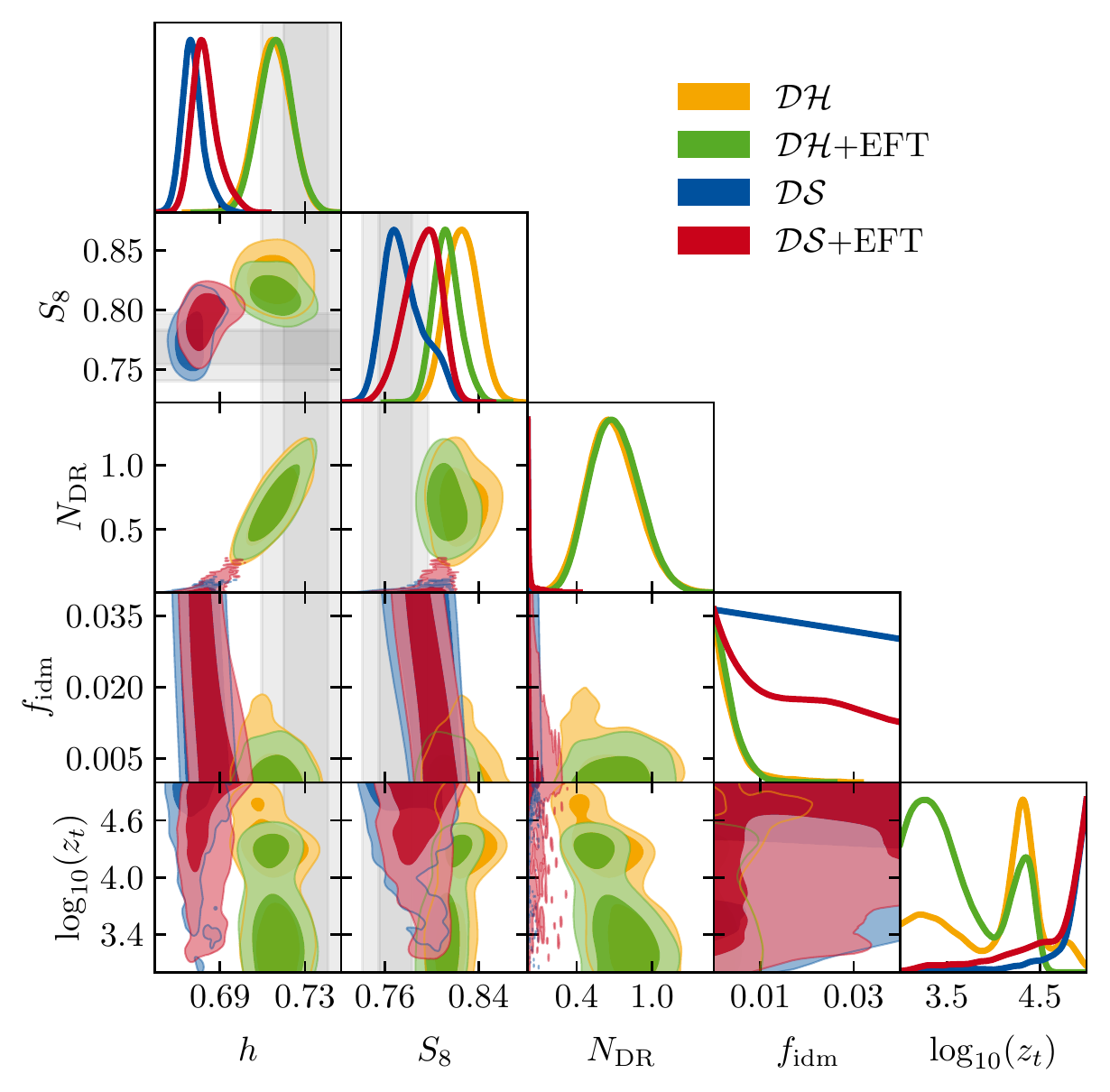}
    \caption{Triangle plot of two-dimensional constraints (68\% and 95\% C.L.) on the parameters of the strongly interacting stepped dark radiation model when confronted with the $\mathcal{DH}$+EFT and $\mathcal{DS}$+EFT data sets.}
    \label{fig:eft_strong}
\end{figure}

\begin{table}
    \centering
    \begin{tabular}{c | c  c}
        Quantity & $\mathcal{DH}$+EFT & $\mathcal{DS}$+EFT \\ \hline 
        \rule{0pt}{3ex} $S_8$ & $0.8061\pm 0.0094$ & $0.794\pm 0.014$ \\
        Shift & $-1.1\sigma$ & $+0.5\sigma$ \\
        Error bar reduction & $40\%$ & $40\%$ \\
        Remaining Gaussian tension & $2.0\sigma$ & $1.2\sigma$
    \end{tabular}
    \caption{Summary of the shifts in $S_8$ for the strongly interacting model induced through the EFTofBOSS and EFTofeBOSS data. The row \enquote{Shift} denotes the shift (in units of the original error bar) of the constraint in $S_8$ compared to the case without EFT. The row \enquote{Error bar reduction} instead denotes the corresponding reduction in the uncertainty.}
    \label{tab:strong_shifts}
\end{table}

For the strongly interacting model, the addition of the EFT analysis does not significantly change the cosmological constraints for the data set $\mathcal{D}$ ($N_\mathrm{DR}< 0.52$ and $f_\mathrm{idm} < 0.027$), as well as for the data set $\mathcal{DHS}$, which is still reduced to the non-interacting stepped dark radiation model (with $N_\mathrm{DR} = 0.78 \pm 0.18$ and $f_\mathrm{idm} < 0.0096$). 
Indeed, adding the EFT for these two data sets produces only a small shift of $0.2\sigma$ on $H_0$ and $-0.5\sigma$ on $S_8$, respectively.
However, if we add the EFT analysis to data sets $\mathcal{DH}$ or $\mathcal{DS}$, we obtain:
\begin{align*}
&N_\mathrm{DR} = 0.78 \pm 0.20 \ {\rm and} \ f_\mathrm{idm} < 0.0098 \quad {\rm for} \ \mathcal{DH}{\rm +EFT},\\
&N_\mathrm{DR}< 0.20 \ {\rm and} \ f_\mathrm{idm} < 0.082 \quad\quad\quad\quad\ \  {\rm for} \ \mathcal{DS}{\rm +EFT},
\end{align*}
which corresponds to a significant improvement on the $f_\mathrm{idm}$ constraints by a factor of 1.5 and 37 with respect to $\mathcal{DH}$ and $\mathcal{DS}$, as can be seen in \cref{fig:eft_strong}.
This improvement of the constraints on $f_\mathrm{idm}$ considerably changes the constraints on $S_8$, as summarized in \cref{tab:strong_shifts}.

In summary, both models are pulled to $S_8 \simeq 0.8$ which is favored by the EFT data. This is most notable for the $\mathcal{DHS}$ and $\mathcal{DS}$ cases in the weakly interacting model, and for the $\mathcal{DH}$ and $\mathcal{DS}$ cases in the strongly interacting model.

\section{Model variations}\label{sec:alt_priors}
In this Section, we investigate variations of the original weakly and strongly interacting models introduced above.

As a first check, we investigate if the performance of the strongly interacting model can be improved when releasing the prior on $\log_{10}(z_t)$ that was motivated by the requirement to keep the dark radiation self-coupled until at least matter-radiation equality (see \cref{app:selfcouple}). We release this requirement under the assumption that some other mechanism keeps the dark radiation self-coupled.

\begin{figure}
    \centering
    \includegraphics[width=\columnwidth]{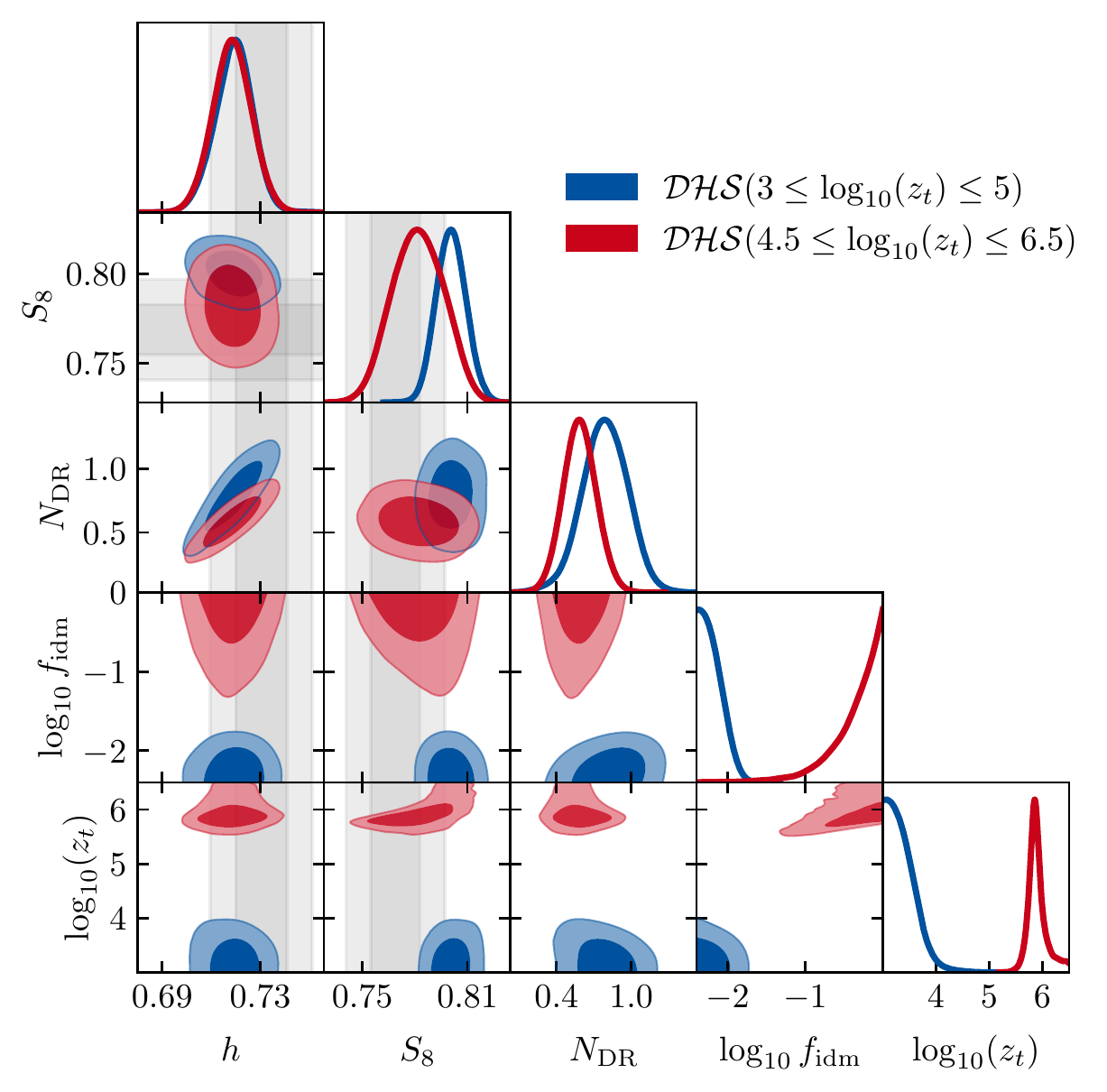}
    \caption{The strongly interacting model is able to resolve both tensions if we allow $z_t$ to exceed its theoretically allowed range. We also show the original run with theoretically motivated prior $\log_{10}(z_t) \in [3,5]$ in blue for comparison.}
    \label{fig:strong_large_zt}
\end{figure}

The constraints on this extended model are shown in \cref{fig:strong_large_zt}, which clearly demonstrates that for \mbox{$\log_{10}(z_t) \simeq 6$} the strongly interacting dark matter model can ease both the Hubble and the $S_8$ tension in this regime (reaching a better fit by $\Delta \chi^2 = -2.81$). 
In order to reach a lower value of $S_8=0.782 \pm 0.015$ (as driven by the $\mathcal{S}$ prior), the model requires a lower limit on $f_\mathrm{idm}> 0.092$ (95\% CL). It is interesting to note that in neither this nor the original prior range there is a preference of values of $\log_{10}(z_t) \sim 4.5$ that are required for the original stepped dark radiation mechanism to ease the Hubble tension. Nevertheless, in both cases the same high value of $H_0 = (71.9 \pm 0.8)$km/s/Mpc can be reached.

\begin{figure}
    \centering
    \includegraphics[width=0.8\columnwidth]{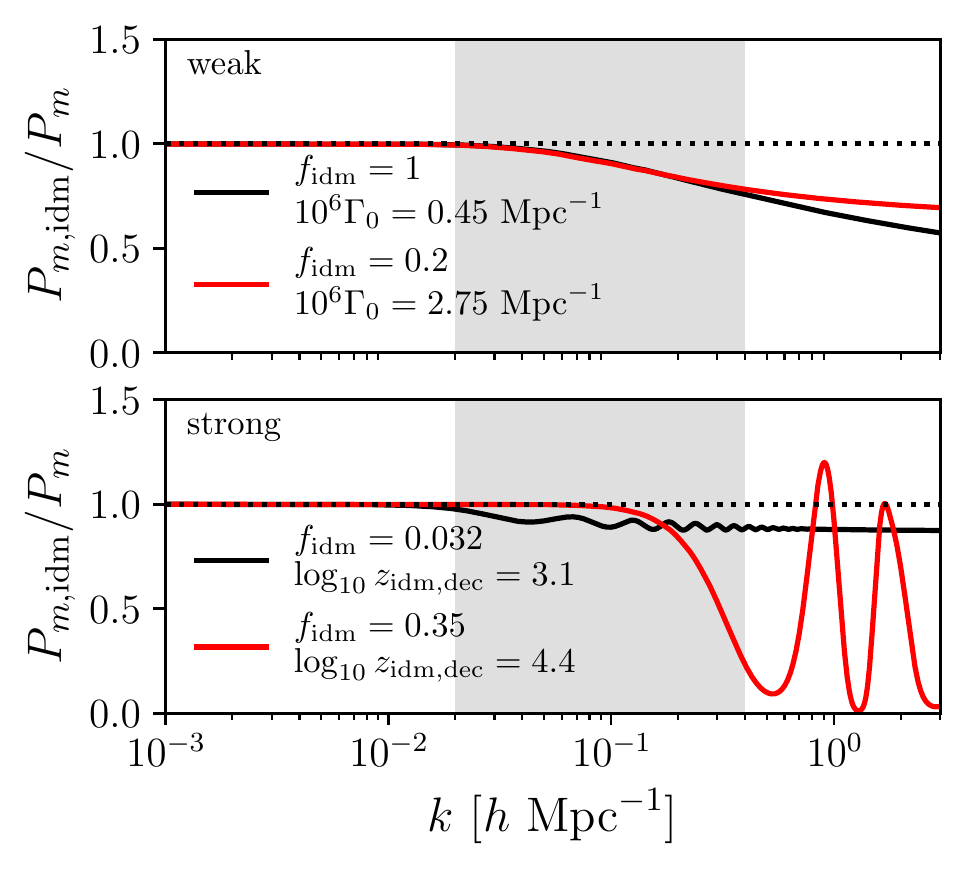}
    \caption{A comparison between the matter power spectrum suppression for two of the extended models we considered divided by their non-interacting limits. Each of the matter power spectra give $\sigma_8 = 0.785$. The top panel compares the standard weak model where all of the dark matter is interacting ($f_{\rm idm} = 1$) (black) to the suppression when $f_{\rm idm}<1$ (red). The bottom panel shows a comparison between the strong model with a model where $z_{\rm idm,dec} < z_{\rm eq}$ (black) and one where $z_{\rm idm,dec} > z_{\rm eq}$ (red). The gray bands roughly indicate the wavenumbers that contribute to $\sigma_8$. }
    \label{fig:comp_Pm_ext}
\end{figure}

The larger suppression allowed for such high $\log_{10}(z_t)$ is driven by it starting only at wavenumbers larger than $k \gtrsim 0.2\mathrm{/Mpc}$, which is simply beyond the range of scales to which the CMB is sensitive, while still allowing for suppression of the power spectrum in the regime relevant for the $S_8$ integral. We expect such a solution to be more strongly constrained from Lyman-$\alpha$ data, which are sensitive to such suppressions of the power spectrum at high wavenumbers (see Ref.~\cite{Archidiacono:2019wdp,Hooper:2022byl}). We can also appreciate from \cref{fig:comp_Pm_ext} that this kind of model leads to large dark acoustic oscillations, which other small-scale probes would likely be sensitive to.

\begin{figure}
    \centering
    \includegraphics[width=\columnwidth]{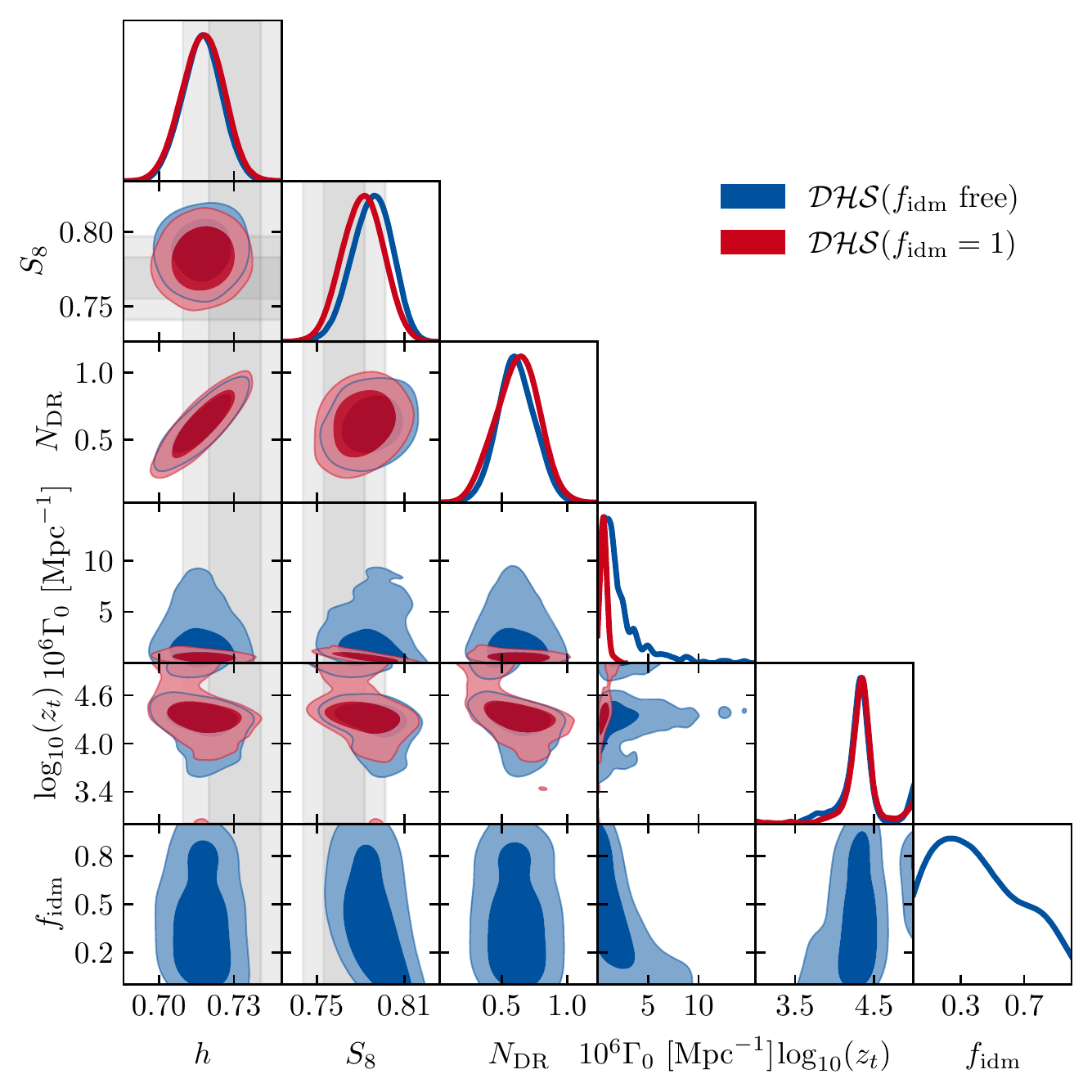}
    \caption{A triangle plot showing the weak model but with only a fraction $f_{\rm idm}$ of the dark matter which interacts with the dark radiation.}
    \label{fig:weak_fidm}
\end{figure}

This sensitivity of Lyman-$\alpha$ data to a large suppression of the matter power spectrum could also impact the weakly interacting model, whose asymptotic suppression goes to 100\%, since all of the dark matter is assumed to be interacting. One way to avoid such strong constraints would be by releasing the fraction of dark matter that interacts in this weakly interacting model. We show the corresponding constraints on the fractional model in \cref{fig:weak_fidm}, with the corresponding power spectrum suppression shown in \cref{fig:comp_Pm_ext} (the asymptote is not reached by $k\sim 2h/\mathrm{Mpc}$). It is worth noting that the constraints on $S_8$ are virtually unchanged, even in the regime of somewhat smaller $f_\mathrm{idm} \sim 0.5$. It is immediately clear from \cref{fig:comp_Pm_ext} that we expect a strong degeneracy between $f_\mathrm{idm}-\Gamma_0$, which we indeed observe in \cref{fig:weak_fidm}. Note that in the $f_\mathrm{idm}-S_8$ panel we can clearly see that for $f_\mathrm{idm} \to 0$ the power spectrum can not be suppressed anymore, leading to larger values of $S_8$ in this limit (as expected). Whether this particular degeneracy of $f_\mathrm{idm}$ and $\Gamma_0$ allows the fractional weakly interacting model to evade Lyman-$\alpha$ constraints remains to be investigated in future work.

\begin{figure}
    \centering
    \includegraphics[width=\columnwidth]{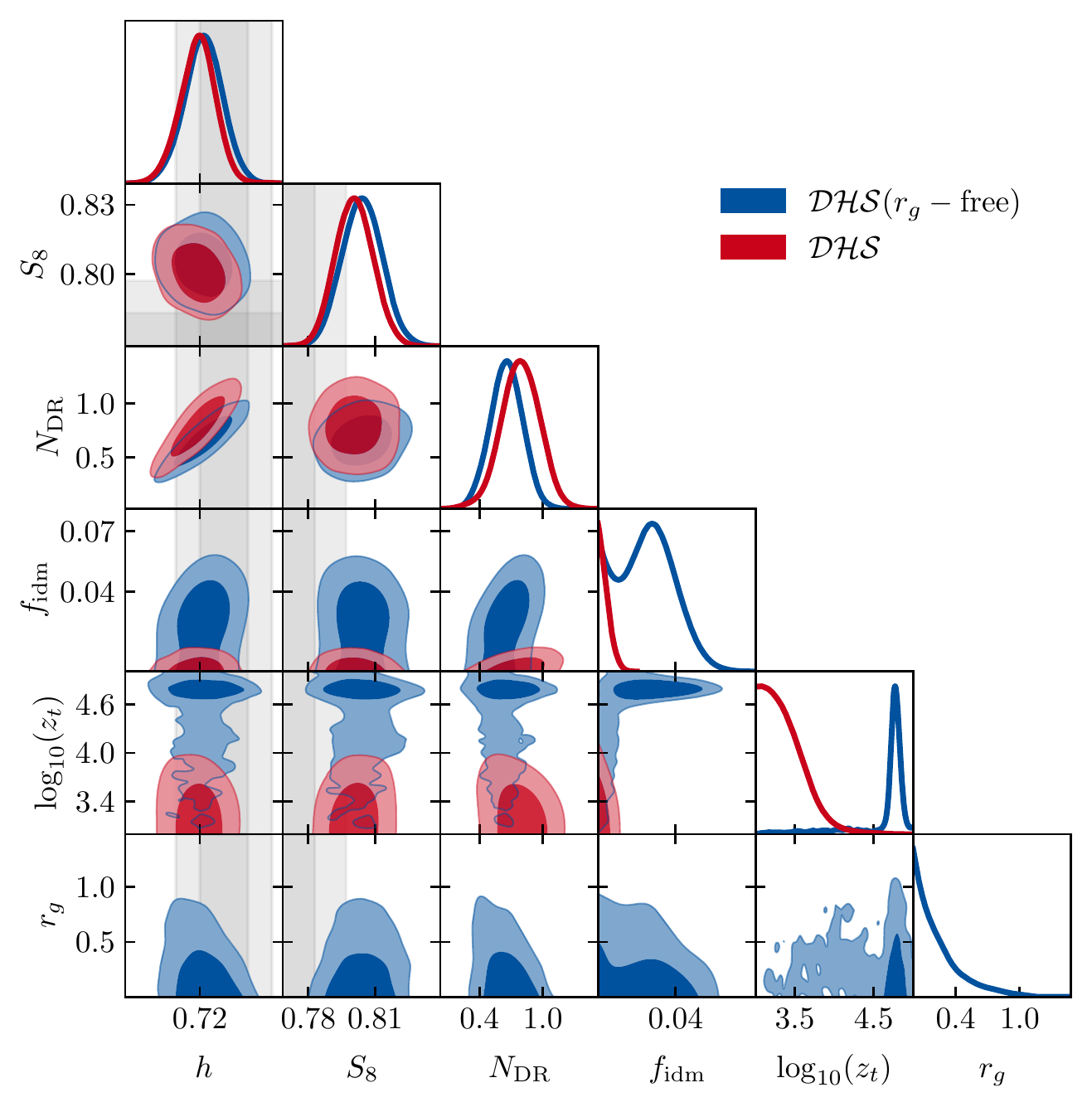}
    \caption{A triangle plot showing the strong model where we now allow the size of the step, $r_g$, to vary.}
    \label{fig:strong_rg_free}
\end{figure}

Finally, we investigate what happens when the step is released for the strongly interacting model. The results, shown in \cref{fig:strong_rg_free}, point to similar abilities of this model in terms of both $H_0$ and $S_8$ compared to the usual model. However, the value of $r_g$ in the original strongly interacting dark radiation model of Ref.~\cite{Buen-Abad:2022kgf} ($r_g = 7/4 = 1.75$) is strongly excluded in this extended parameter range ($r_g< 1.1$ at 95\%CL). Indeed, in this case, the minimal $\chi^2$ is 3816.0 (compared to 3821.6 in the original model). On the other hand, for such a small step, comparatively larger fractions of $f_\mathrm{idm}< 0.049$ are allowed for this model. Otherwise the conclusions remain similar, with $H_0 = (72.15 \pm 0.83)$km/s/Mpc and $S_8={0.8042}^{+0.0088}_{-0.0092}$.

\section{Conclusions}\label{sec:conclusions}
\enlargethispage*{1\baselineskip}
We have shown that the two recently proposed models of interacting stepped dark radiation from \cite{Joseph:2022jsf,Buen-Abad:2022kgf} are both exciting proposals that could simultaneously ease the Hubble and $S_8$ tensions. While the model of \cite{Joseph:2022jsf} is weakly interacting and shows a shallow suppression of the power spectrum, the model of \cite{Buen-Abad:2022kgf} interacts strongly and has a sharp suppression of the power spectrum and dark acoustic oscillations. These oscillations happen to also cause a stronger driving of the CMB source terms, resulting in an overall larger impact on the anisotropies. Since the decoupling of the dark radiation from the dark matter is also delayed in this strongly interacting model, we find that the parameter space in which the Hubble tension is eased better (around $\log_{10}(z_t) \sim 4.5$ in both models) is slightly different to the parameter space where the $S_8$ tension can be addressed without impacting CMB observables (requiring either $\log_{10}(z_t)\lesssim 3.5$ or $\log_{10}(z_t) \gtrsim 4.8$). This reduces the strongly interacting model's ability to ease both tensions simultaneously. We find that nevertheless both models are quite effective.

Without enacting any priors beyond our baseline Planck+BAO+Pantheon dataset $\mathcal{D}$, we find that the Hubble tension remains at the $3\sigma$ level, while the $S_8$ tension is eased at the $2 \sigma$ level in the strongly interacting model and at the $1 \sigma$ level in the weakly interacting model. However, when faced with the task of easing only the $S_8$ tension, both models fair similarly well ($S_8=0.786\pm0.020$ for the strongly interacting model, and $S_8=0.778\pm0.014$ for the weakly interacting model, when an $S_8$ prior is added). Similarly, when imposing only an $H_0$ prior, the Hubble tension is reduced to the $1 \sigma$ level in both models (reaching approximately $(71.6 \pm 0.8)$km/s/Mpc in both models). However, the difference between the models becomes obvious when imposing both priors simultaneously. Here the $S_8$ tension is only reduced to $1.7\sigma$ in the strongly interacting model but to $0.6\sigma$ in the weakly interacting model -- while the Hubble tension remains at the $1\sigma$ level for both models. This manifests in a strong preference compared to $\Lambda$CDM when both of these priors are imposed, with $\Delta$AIC=$-20.5$ for the weakly interacting model and $\Delta$AIC=$-13.8$ for the strongly interacting model.

Given that in the weakly interacting model both tensions are eased in a synergistic way, while for the strongly interacting model the easing of one tension often comes at the cost of easing the the other, we might wonder if additional data can further disambiguate these two mechanisms. As such, we subject both models to ACT+SPT high-$\ell$ polarization data, which has been shown in the past to be quite sensitive to additional dark radiation. We find that even in this case the weakly interacting model remains more synergistic, with an $S_8$ tension of $0.6\sigma$ compared to the strongly interacting model with a tension of $1.8\sigma$. Interestingly, both models exhibit an much better fit to Planck+ACT+SPT data than the $\Lambda$CDM model, with a $\Delta\chi^2_{ACT+SPT+Planck}\lesssim-10$. A similar effect has been noted in the context of early dark energy in Ref.~\cite{Smith:2022hwi}, but a more detailed analysis is left for future work. This effect also results in good overall preferences of $\Delta$AIC=$-27.7$ in the weakly interacting model compared to $\Lambda$CDM, while the strongly interacting model reaches $\Delta$AIC=$-18.4$.

Since both models feature a suppression of the power spectrum, other large scale structure data is of vital importance in constraining these models. In this work we make use of the full-modeling EFTofLSS data from eBOSS DR16 QSOs and BOSS DR12 LRG (together denoted as EFT), and we show that there is a strong impact on both models. We have also checked that $f\sigma_8$ data (from redshift space distortions) is insensitive to these suppressions and does not give the same constraining power as full-modeling.
In the weakly interacting model the effect is a much tighter constraint on $\Gamma_0$ by a factor of 3.3 and an corresponding increase of the $S_8$ tension to $1.4\sigma$ (with our baseline data, the $H_0$ and $S_8$ priors, and the EFT data). In the strongly interacting model the effect is only a mild increase in constraining power when the EFT data is considered in addition to the baseline data and the priors, since already without the EFT data the model is tightly constrained. In this case the power of the EFT data is most appreciable when considering a combination of either only the Hubble prior or only the $S_8$ prior, the latter featuring a factor of 37 in improvement of constraining power on the fraction of interacting dark matter, $f_\mathrm{idm}$\,.
In summary, we find that the EFT data seems to constrain both models to values of $S_8$ closer to about $0.8$, and correspondingly restricts their ability to ease the $S_8$ tension (while leaving the mechanism of easing the Hubble tension largely unchanged), with both models featuring a $\sim 1.5\sigma$ tension in $S_8$ when the baseline data, the two priors, and EFT data are added.

We have also checked if certain assumptions that go into the model-building of the two models can be released in order to find more compelling solutions. While we put an upper prior on the transition redshift $z_t$ for the strongly interacting model motivated from a requirement to keep the dark radiation self-coupled, if we release this prior (assuming some other mechanism to self-couple the dark radiation), we find that another compelling solution to the tensions emerges in a regime of $z_t \sim 10^6$. This solution does reach $S_8 = 0.782\pm 0.015$ ($0.6\sigma$ tension) even with the Hubble prior (and $S_8$ prior) imposed.

Another possible avenue to increase the ability of the strongly interacting model's ability to ease both tensions is to release its theoretically motivated relative abundance of dark radiation before and after the step, characterized by the step parameter $r_g$\,. Such a model does improve the overall fit by $\Delta \chi^2 = -5.6$ (enough to justify the additional parameter with respect to the AIC criterion). However, we find that this variation does not significantly impact the model's ability to ease the tension, with $S_8$ decreasing only marginally, at the cost of excluding the original $r_g=1.75$ by more than $2 \sigma$ (we have $r_g < 1.1$ at 95\% CL).

All of these models and their respective variations significantly suppress the power spectrum on non-linear scales around $k \gtrsim 0.5h$/Mpc. We thus expect other data probing these smaller scales to put significant constraints on such a model. In particular, Lyman-$\alpha$ forest data can measure these scales at high redshift when the non-linearities of structure formation are not quite as pronounced. While a systematic study of the impact on Lyman-$\alpha$ data is beyond this work\footnote{A simple treatment with an amplitude and a slope at a given scale and redshift such as in would certainly be possible, but it has not yet been proven that such a treatment as in \cite{Pedersen:2022anu,Goldstein:2023gnw} is valid even in cosmologies with large suppressions of the power spectrum at small scales. As such, we leave this study to future work.}, we can already estimate that the $z_t$-extended strongly interacting model and possibly the weakly interacting model will be impacted by such Lyman-$\alpha$ data due to their large (and growing) suppression at scales $\gtrsim 1h$/Mpc. This motivates us also to search for a weakly interacting model in which only a fraction of the dark matter interacts, leading to a strong $\Gamma_0-f_\mathrm{idm}$ degeneracy, which Lyman-$\alpha$ data is poised to further constrain (such as in \cite{Archidiacono:2019wdp}).

Beyond Lyman-$\alpha$ data we expect other small-scale data such as the upcoming DESI and Euclid galaxy and weak lensing surveys to tightly increase the constraint on such interacting dark radiation models. Similarly, upcoming CMB observations of the large-$\ell$ polarization will uniquely distinguish stepped dark radiation models and allow for precise statements on their viability. There is thus little doubt that these interacting stepped dark radiation models are not only exciting in terms of their current ability to ease the Hubble and $S_8$ tensions, but also that upcoming constraints are going to decisively rule in favor or against these models.

\vspace*{\baselineskip} 
\noindent
\textbf{Note added}: During the final stages in the preparation of this manuscript, Ref.~\cite{Allali:2023zbi} and Ref.~\cite{Buen-Abad:2023uva} appeared, which have tested interacting dark sector models using cosmological data. On the one hand, Ref.~\cite{Allali:2023zbi} considered both weak and strong models presented here, but have treated the step size $r_g$, the interacting dark matter fraction $f_{\rm idm}$ and the scattering rate $\Gamma_0$ as free parameters in both models. This reference finds more pessimistic results in terms of the $H_0$ and $S_8$ tensions for the two models, mainly due to the different number of free parameters and the priors chosen (i.e., the use of logarithmic priors on $\Gamma_0$ and $f_{\rm idm}$, as opposed to linear). On the other hand, Ref.~\cite{Buen-Abad:2023uva} considered just the strong model as well as a theoretically-motivated generalization of it with a smaller step size $r_g$. This reference finds a preference for this smaller stepsize as a solution to both tensions, in good agreement with our discussion at the end of \cref{sec:alt_priors}. Let us also emphasize that none of these works have considered small-scale CMB measurements from ACT and SPT, nor eBOSS data analyzed under the EFTofLSS, as we do. Furthermore, our work is the first to provide a detailed comparison between the impact of the weakly and strongly interacting models on the CMB spectra.

\begin{acknowledgements}
N.~S.~acknowledges support from the Maria de Maetzu fellowship grant: CEX2019-000918-M, financiado por MCIN/AEI/10.13039/501100011033. G.~F.~A. is supported by the European Research Council (ERC) under the European Union's Horizon 2020 research and innovation programme (Grant agreement No. 864035 - Undark). T.~S.~acknowledges the use of computational resources from the LUPM's cloud computing infrastructure founded by Ocevu labex, and France-Grilles. A.~B.~,Y.~P.~, and T.~L.~S.~are supported by Grant No.~AST-2009377. This work used the Strelka Computing Cluster, which is run by Swarthmore College. T.~L.~S.~thanks Westwinds Farm where part of this work was completed. This project has received support from the European Union’s Horizon 2020 research and innovation program under the Marie Skodowska-Curie grant agreement No 860881-HIDDeN. This project has also received funding from the European Research Council (ERC) under the European Union’s HORIZON-ERC-2022 (Grant agreement No. 101076865).
\end{acknowledgements}

\bibliography{bibliography}

\begin{appendix}

\section{Adiabatic initial conditions}\label{app:initial}

It is important to realize that the initial conditions (ICs) of the interacting stepped dark radiation models are not entirely trivial. In this Section, we will derive the adiabatic initial conditions for both the dark radiation and interacting dark matter. We find and implement slightly different ICs compared to what is implemented in \texttt{CLASS v3.2} for interacting dark radiation, although the issue is known and will be fixed in future versions of \texttt{CLASS} (internal communication with the \texttt{CLASS} developers). We note that in the two models we consider here, using the correct ICs leads to a negligible change ($\sim 10^{-3}$) in the resulting power spectra. 

We follow the derivation of adiabatic ICs first presented in Ref.~\cite{Ma:1995ey}. The starting point is the fact that we set the initial conditions during radiation domination, so the only sources for the gravitational potentials will be due to the radiative components. In synchronous gauge, this means that 
\begin{equation}
\eta^2 \ddot h + \eta \dot h + 6[R_\gamma \delta_{\gamma} + R_\nu \delta_\nu + R_{\rm DR} \delta_{\rm DR}]=0,
\end{equation}
where $R_X \equiv \rho_X/\rho_{\rm rad}$ is the fractional contribution from each radiative component with respect to the total radiative energy density; with this definition, we have $R_\gamma = (1-R_\nu- R_{\rm DR})$. 
In addition to this, the fluid equations immediately tell us that to lowest order in $k\eta$ we can neglect the velocity perturbations, $\theta$, compared to the density contrasts $\delta$. This gives 
\begin{equation}
\delta_b = \delta_c = \delta_{\rm idm} = \frac{3}{4} \delta_{\gamma}=\delta_{\nu}=\delta_{\rm DR} = -\frac{2}{3} h,
\end{equation}
which allows us to find (keeping only the growing solution)
\begin{eqnarray}
h &=& C (k\eta)^2,\\ 
\delta_\gamma &=& \delta_\nu = \delta_{\rm DR} = -\frac{2}{3} C (k \eta)^2. \nonumber
\end{eqnarray}

Furthermore, the time-time Einstein equation can then be re-written as
\begin{eqnarray}
k^2 \eta_S - \frac{\dot h}{2\eta} &=& -\frac{3}{2}\frac{1}{\eta^2} (R_\gamma \delta_\gamma + R_\nu \delta_\nu + R_{\rm DR} \delta_{\rm DR})\\ &=& -\frac{3C}{2} k^2,\nonumber
\end{eqnarray}
which immediately gives, to leading order, 
\begin{equation}
\eta_S \approx 2C~.
\end{equation}
In \texttt{CLASS} we set $C = 1/2$ for the adiabatic initial conditions (which corresponds to normalizing with respect to the comoving curvature perturbation).

The ICs are set while the baryons and photons are tightly coupled so that the photon anisotropic stress vanishes at leading order and $\theta_b = \theta_\gamma$. This gives 
\begin{equation}
\dot \theta_{\gamma} - \frac{1}{4} k^2 \delta_{\gamma} = 0,\label{eq:theta_ini_gamma}
\end{equation}
which can be easily solved to give 
\begin{equation}
\theta_\gamma  = - \frac{1}{18} C(k^4 \eta^3).
\end{equation}
\begin{figure}
    \centering
    \includegraphics[width=\columnwidth]{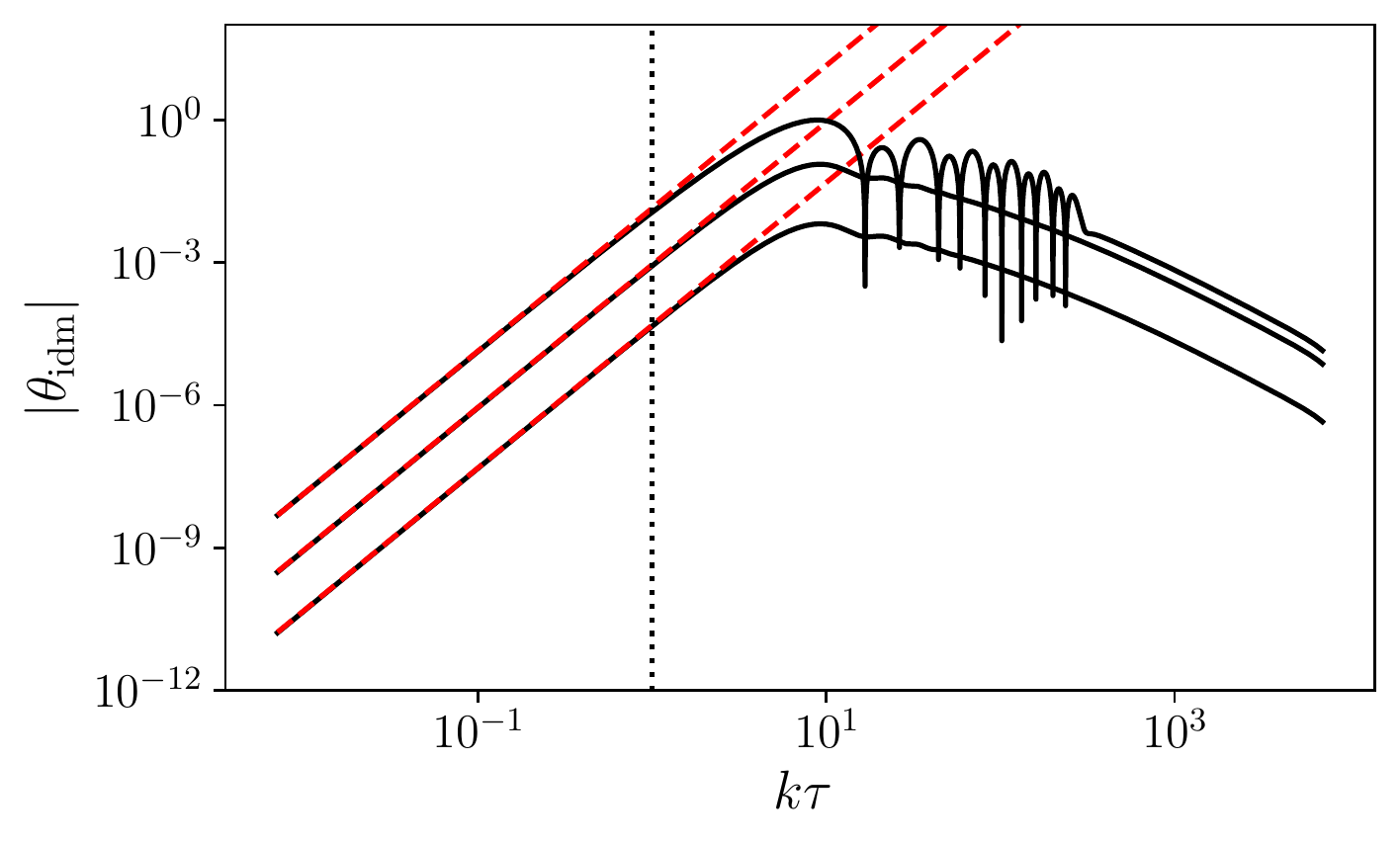}
    \caption{The super and sub horizon evolution of the interacting dark matter in the weakly interacting model. The dashed red curves show the initial conditions in \cref{app:eq:ic_idm}. The top curves correspond to $\Gamma_0 = 1\ {\rm Mpc}^{-1}$ ($[\Gamma/H]_{\rm init} = 2.8 \times 10^5$), the middle curves correspond to $\Gamma_0 = 10^{-6}\ {\rm Mpc}^{-1}$ ($[\Gamma/H]_{\rm init} = 0.28$), and the bottom curves correspond to $\Gamma_0 = 5 \times 10^{-7}\ {\rm Mpc}^{-1}$ ($[\Gamma/H]_{\rm init} = 0.014$). The agreement between the dashed-red and solid-black curves shows that our initial conditions apply to both initial strong and weak interactions. To produce this figure we set all $\Lambda$CDM parameters to their best fit values to \textit{Planck} and $N_\mathrm{DR} = 0.5$, $\log_{10}(z_t)= 4.5$. }
    \label{fig:theta_idm}
\end{figure}
Finally, we need to account for the momentum exchange between the interacting dark matter and dark radiation. Writing out \cref{eq:euler_idm} we get
\begin{equation}\label{app:eq:euler_idm}
    \dot{\theta}_{\rm idm} = - \mathcal{H} \theta_{\rm idm} + a \Gamma (\theta_{\rm DR}- \theta_{\rm idm}),\\
\end{equation}
which can be directly integrated to give 
\begin{equation}\label{app:eq:integrated_idm}
    \theta_{\rm idm} = A \eta^{-(\kappa+1)} + \kappa \left[\int x^\kappa  \theta_{\rm DR}(x) dx\right] \eta^{-(\kappa+1)}~,
\end{equation}
with $\kappa \equiv \Gamma/H$ and $A$ being an irrelevant integration constant for a decaying solution. Here we assumed that $\kappa = \mathrm{const}$ during the setting of the initial conditions, which we enforce by setting the initial conditions always before $z_t$\,. In this regime, we can also simplify the dark radiation \cref{eq:euler_wzdr} as
\begin{equation}\label{app:eq:euler_wzdr}
    \dot{\theta_{\rm DR}} = - \frac{k^2}{3} \delta_\mathrm{DR} + a \Gamma S (\theta_{\rm idm}- \theta_{\rm DR})~,
\end{equation}
where $S \equiv \rho_{\rm idm}/(\rho_{\rm DR}+P_{\rm DR})$. It is straightforward to show from \cref{app:eq:euler_wzdr} that as long as $\Gamma S/H \ll 1$, to leading order in $k\eta$ we have 
\begin{equation}\label{app:eq:ic_wzdr}
\theta_{\rm DR} = -\frac{C}{18}k^4 \eta^3,
\end{equation}
and from \cref{app:eq:euler_idm} (or even faster using \cref{app:eq:integrated_idm}) we then get
\begin{equation}\label{app:eq:ic_idm}
\theta_{\rm idm} = -\frac{C \kappa}{18 (4 + \kappa)}k^4 \eta^3.
\end{equation} 
We can see that in the limit of strong coupling ($\kappa \gg 1$) we have $\theta_{\rm idm} \approx \theta_{\rm dr}$, and even for weak coupling the interacting dark matter has a non-zero initial velocity perturbation. Note that in the weakly interacting model we have $[\Gamma/H]_{\rm init} \approx \Gamma_0 (T_\mathrm{DR} a/T_\mathrm{DR,0})^2/(H_0 \sqrt{\Omega_r}) = \Gamma_0 (7/15)^{2/3}/(H_0 \sqrt{\Omega_r})$, see Ref.~\cite{Joseph:2022jsf}.

In \cref{fig:theta_idm} we show a comparison between the analytic initial condition for $\theta_{\rm idm}$, \cref{app:eq:ic_idm}, in dashed red, and the output from our modified version of \texttt{CLASS}, in solid black, for $k=0.5\ {\rm Mpc}^{-1}$\,. The three pairs of curves correspond to different values for the interaction rate, $\Gamma_0 = 1\ {\rm Mpc}^{-1}$  (upper curves), $\Gamma_0 = 10^{-6}\ {\rm Mpc}^{-1}$ (middle curves) and $\Gamma_0 = 5 \times 10^{-8}\ {\rm Mpc}^{-1}$ (lower curves)-- which give $[\Gamma/H]_{\rm init} = 2.8 \times 10^5$, $[\Gamma/H]_{\rm init} = 0.28$, and $[\Gamma/H]_{\rm init} =0.014 $, respectively. The vertical dotted line shows horizon crossing. Here we have plotted the weakly interacting model, while the strongly interacting results are nearly identical (differing mostly through $N_\mathrm{DR}(z=0)/N_\mathrm{DR}(z\to \infty)$ and the different interaction rate penalty factor for $T \ll m$, both of which are only relevant at around $z_t$ and lower, not affecting the initial conditions). 

In the strongly interacting case, we have $\Gamma/H \gg 1$ at initial times, which means that $S \Gamma/H \ll 1$ is in principle not entirely trivial to enforce. However, by inserting the limit of $\mathcal{\kappa} \gg 1$ in \cref{app:eq:integrated_idm} one can show that in this case only $S \Gamma/H \cdot 1/\kappa \ll 1$ is required, which is equivalent to the weaker condition $S\ll1$ by using $\kappa = \Gamma/H$. In the weakly interacting case, we have $\Gamma/H \sim 1$ and $S \ll 1$ is sufficient. In practice, we simply enforce the initialization scale factor to be early enough for $S \ll 1$ to be fulfilled.

We cannot immediately solve the neutrino Euler equation for its velocity perturbation because of the unknown neutrino anisotropic stress. The time-space component of the Einstein equation allows us to relate $\eta_S$ and $\theta_\nu$: 
\begin{eqnarray}
\dot \eta_S &=& \frac{8}{9(k \eta)^2}(R_\gamma \theta_\gamma + R_{\nu} \theta_\nu + R_{\rm DR} \theta_{\rm DR}),\\
&=& \frac{8}{9(k \eta)^2}\left[\theta_\gamma (1-R_\nu) + R_\nu \theta_\nu\right],
\end{eqnarray}
where in the second line we used the fact that we know $\theta_{\rm DR}=\theta_\gamma$. Finally, we have the neutrino Euler equation and equation for the anisotropic stress
\begin{eqnarray}
\dot \theta_\nu - \frac{1}{4} k^2(\delta_\nu - 4 \sigma_\nu) &=& 0,\\
\dot \sigma_\nu - \frac{2}{15} (2 \theta_\nu + \dot h + 6\dot \eta_S) &=& 0.
\end{eqnarray}
We now have three equations for the three unknown initial conditions ($\eta_S$, $\theta_\nu$, and $\sigma_\nu$) and we can solve them to leading order and find 
\begin{eqnarray}
\eta_S &=& 2 C - \frac{5+4 R_\nu}{6(15+4 R_\nu)}C (k\eta)^2,\\
\theta_\nu &=& \frac{23+4 R_\nu}{15+4 R_\nu} \theta_\gamma,\\
\sigma_\nu &=& \frac{4C}{3(15+4 R_\nu)}(k\eta)^2.
\end{eqnarray}

In summary, we showed that as long as we enforce $S(z_\mathrm{ini}) \ll 1$ and $z_\mathrm{ini} > z_t$ during the setting of the initial conditions, then our modified initial conditions of \cref{app:eq:ic_wzdr,app:eq:ic_idm} are valid (and the other $\Lambda$CDM initial conditions as well).

\section{Limits of the modeling}\label{app:limits_SD}
While we attempted to keep our modeling of the two models very general, there are a few regions in parameter space where it is not guaranteed that our modeling is complete. One particular concern is the assumption of the dark radiation remaining self-interacting in the strongly interacting model and not being subject to strong spectral distortions (or even deviating from the $T_\mathrm{DR} \propto (1+z)$ scaling). For the energy dissipation, it is relatively straightforward to estimate the size of the spectral distortions due to the fact that the interactions with the dark matter are of the order \footnote{See the related discussion in \cite{Ali-Haimoud:2021lka,Ali-Haimoud:2015pwa}, which focuses on interaction with electrons/baryons/photons, but can easily be generalized to dark radiation interactions.}
\begin{equation}
    \frac{\Delta \rho_\mathrm{DR}}{\rho_\mathrm{DR}} \approx -\frac{3}{2} \int_{-\infty}^{0} r_{\chi,\mathrm{DR}} \frac{(\rho_\mathrm{idm}/m_\mathrm{\chi}) T_\mathrm{DR}}{\rho_\mathrm{DR}} d \ln a~,
\end{equation}
where $r_{\chi,\mathrm{DR}}$ is a coupling coefficient with $r_{\chi,\mathrm{DR}} \to 1$ during tight coupling (defined as in \cite{Ali-Haimoud:2021lka,Ali-Haimoud:2015pwa}), otherwise it is strictly smaller, and $m_\chi$ is the interacting dark matter mass. To derive an estimate of the upper bound, we can simply assume the species to be tightly coupled throughout the evolution and use the simple scaling laws of $\rho_\mathrm{DR} = g_\mathrm{DR} \pi^2/30 T_\mathrm{DR}^4$ (with a possible factor of $7/8$ for fermionic dark radiation) and $\rho_\mathrm{DR}(z=0)=\rho_\gamma \cdot \frac{7}{8} \left(\frac{4}{11}\right)^{4/3} N_\mathrm{DR}$ to find 
\begin{equation}
    \label{app:eq:Tdr}
    T_\mathrm{DR}(z) = \left(\frac{14}{8g_\mathrm{DR}}\right)^{1/4} \left(\frac{4}{11}\right)^{1/3} [N_\mathrm{DR}(z)]^{1/4} \cdot  T_\gamma(z)~,
\end{equation}
and correspondingly
\begin{align*}
    &\frac{N_\chi T_\mathrm{DR}(z)}{\rho_\mathrm{DR}(z)} \approx 1.95 \cdot 10^{-10} f_\mathrm{idm} \relpow{\Omega_\mathrm{cdm} h}{0.12}{}   \relpow{100\mathrm{GeV}}{m_\chi}{} \\
    &\times \relpow{N_\mathrm{DR}}{0.1}{-3/4} \relpow{T_{\mathrm{DR},0}}{a T_\mathrm{DR}(z)}{3} \relpow{g_\mathrm{DR}}{2}{-1/4} \relpow{T_\gamma}{2.72\mathrm{K}}{-3/4}~.
\end{align*}
This implies that over a given decade in redshift the  relative amount of energy extracted from the dark radiation due to its dark matter scattering is tiny as long as the values of all parameters are reasonably close to their fiducial values. The only issues can occur when $N_\mathrm{DR}$ becomes very small. Only as $N_\mathrm{DR}$ approaches the order of $\sim 10^{-12}$ can this effect -- over the complete evolution in the code (32 $e$-folds in scale factor) -- introduce an order unity change of energy density. We thus conclude that for the parameter range considered in this work, we do not need to worry about modifying the temperature equations. As almost all points sampled in the MCMC will also have approximately $N_\mathrm{DR} \gtrsim 10^{-5}$, we will also not worry about possible spectral distortions of the interacting stepped dark radiation models.

\section{Self-interacting decoupling of the dark radiation}\label{app:selfcouple}

From dimensional analysis, one expects a self-interaction rate of the dark photon in the spartacous model of size (c.f. \cite[Eq. (3.5)]{Buen-Abad:2022kgf})
\begin{equation}
    \Gamma_\mathrm{self} = C_\mathrm{self} \frac{\alpha_d^4}{m^8} T_\mathrm{DR}^9~,
\end{equation}
where $C_\mathrm{self}$ is some unknown order-one constant. We can then simply find the redshift of $\Gamma_\mathrm{self} =H$ at which the dark radiation self-interaction will become inefficient, which gives us the condition 
\begin{align*}
    1+z &= \relpow{H(z)}{(1+z)^{3/2}}{2/15} \relpow{a T_\mathrm{DR}(z)}{T_\mathrm{DR,0}}{-6/5} (1+z_t)^{16/15} \\
    &\times \alpha_d^{-8/15}~ T_\mathrm{DR,0}^{-2/15}~ C_\mathrm{self}^{-1}~.
\end{align*}
Here we have used $m = T_\mathrm{DR,0} \cdot (1+z_t)$ to eliminate the mass. We can further simplify many of these terms in order to get a simple expression under a few assumptions. First, we can use \cref{app:eq:Tdr} at $z=0$ to relate $T_\mathrm{DR,0}$ to known CMB quantities such as $T_\mathrm{CMB}$ as well as the model parameter $N_\mathrm{DR}$ and $g_\mathrm{DR}$, which we set simply to 2 for this case. We further approximate $H(z) \approx \sqrt{\Omega_m h^2} H_0/h (1+z)^{3/2}$ during the time of interest (we assume matter domination), and we approximate $a T_\mathrm{DR} \approx T_\mathrm{DR,0}$ (which is justified since the decoupling is typically far from the step, where the usual $T \propto a^{-1}$ scaling applies). We further use $\alpha_d = 10^{-4}$ \,, some approximate value of $\Omega_m h^2 \approx 0.14$\,, and $N_\mathrm{DR}^{-1/30} \approx 1$ (for any reasonable values of $N_\mathrm{DR}$) to find the estimate
\begin{equation}
    1+z_\mathrm{self-dec} \approx 0.017 C_\mathrm{self}^{-1} \cdot (1+z_t)^{16/15}~.
\end{equation}
As long as $C_\mathrm{self}$ is close to unity, and $1+z_t$ is not too large, the self-decoupling of the dark radiation happens around or after recombination, thus justifying the fully self-coupled treatment. With $z_t = 10^5$ (our upper bound), this computes to a decoupling just around radiation-matter equality with $z_\mathrm{self-dec} \sim 3 000$, which should not significantly affect most observables. One could either increase $\alpha_d$ or decrease the upper prior bound to be even more secure of a fully self-coupled dark radiation, and we could conduct a more detailed investigation of this point for future work. 

\section{Cosmological impact}\label{app:impact}

\begin{figure*}[t]
    \centering
    \includegraphics[width=0.4\textwidth]{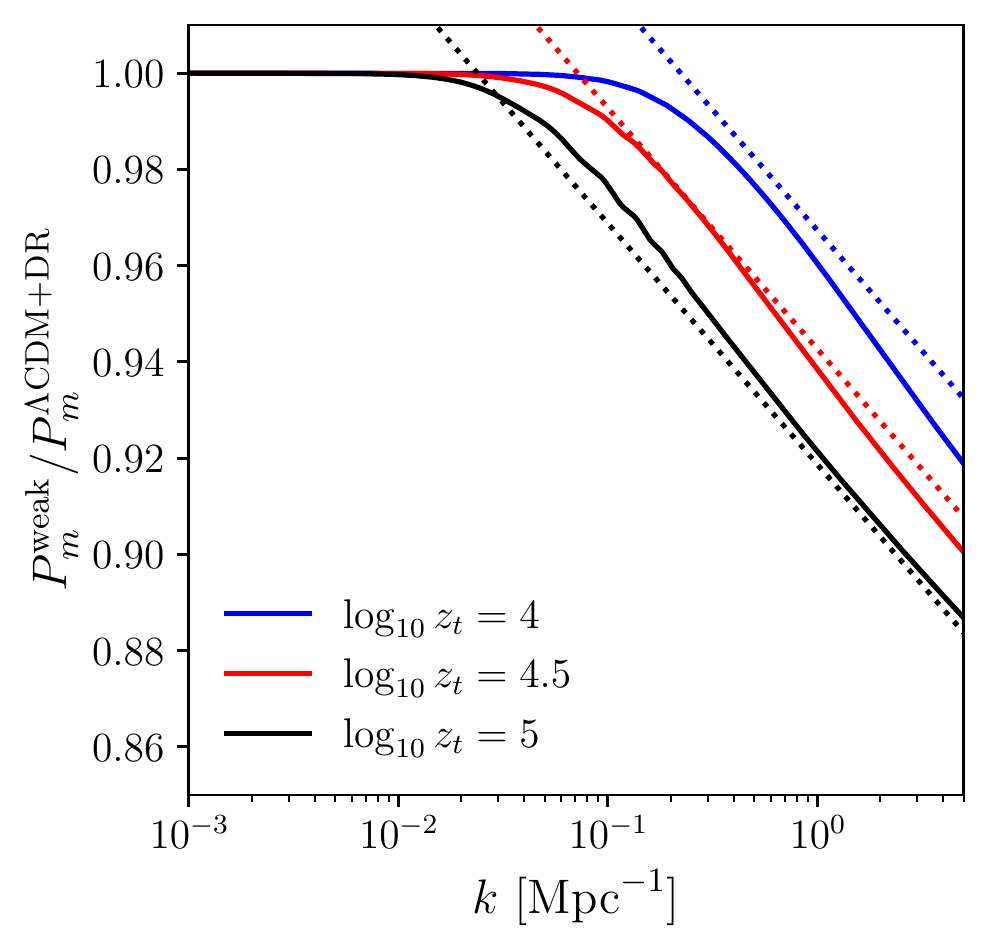}
    \includegraphics[width=0.4\textwidth]{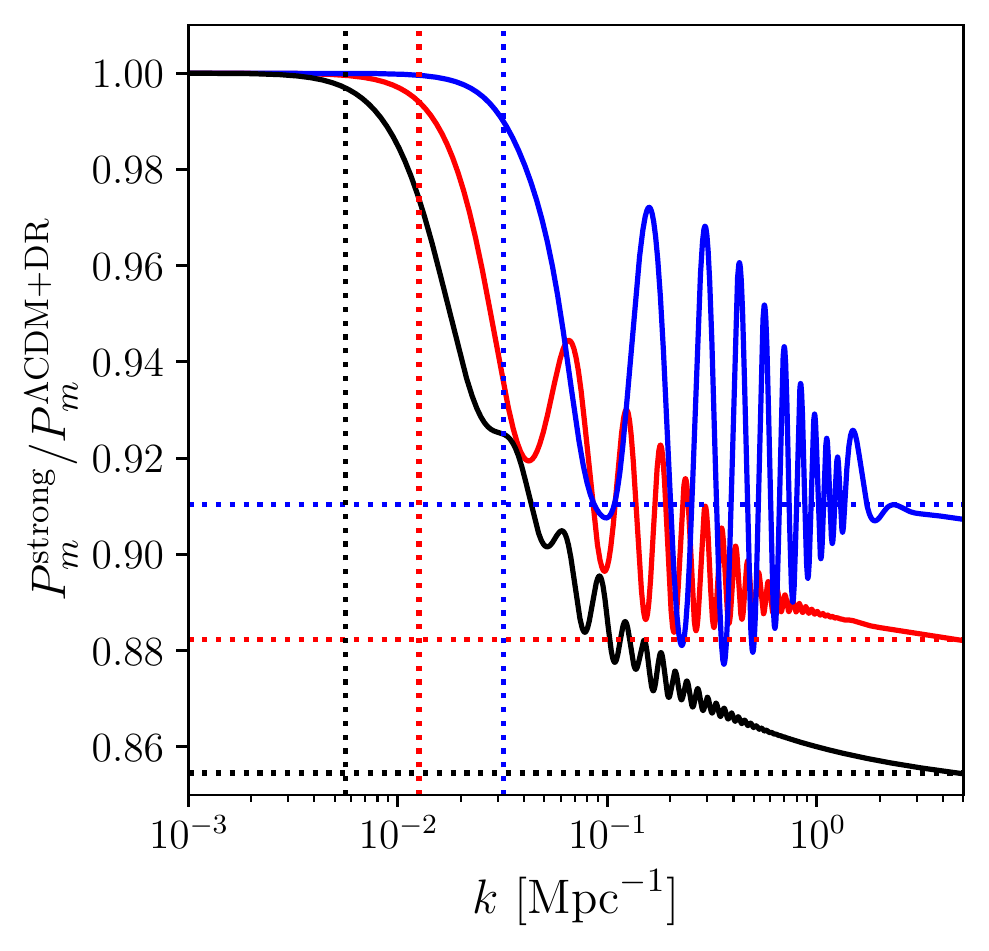}
    \caption{The impact of both the weak (\emph{left}) and strong (\emph{right}) models on the matter power spectrum. The different solid curves correspond to different values of $\log_{10}(z_t)$: 4 (black), 4.5 (red), 5 (blue). The dotted curves in the left panel show the approximate logarithmic suppression for the weak case, \cref{app:eq:Pm_weak} and the horizontal dotted lines in the right panel show the approximate asymptotic suppression, \cref{app:eq:Pm_strong}, in the strong case. The vertical dotted lines in the right panel show the approximate scale at which the suppression starts in the strong case: $k_c \simeq \pi/\tau_{\rm idm,dec}$. For both cases we chose $\Delta N_{\rm DR} = 0.5$; for the weak case we used $\Gamma_0 = 8 \times 10^{-8}\ {\rm Mpc}^{-1}$ (which corresponds to $\kappa = 0.02$) and for the strong case $f_{\rm idm} = 0.03$. Note that the power spectra are normalized by a stepped radiation model (without interacting dark matter). For the strong model the values of $z_t$ correspond to $\log_{10}(z_{\rm idm,rec}) = 2.6, 3.1, 3.6$, respectively.}
    \label{fig:comp_Pm}
\end{figure*}

The different dynamics in the two models can be understood in a simple way. First, in the weakly interacting model, all of the dark matter weakly interacts with the dark radiation. This means that soon after horizon entry, a given mode follows an approximate equation of motion (written in conformal Newtonian gauge)
\begin{equation}
    \ddot{\delta}_{\rm idm}+\mathcal{H}\left(1+\frac{\kappa}{\sqrt{1+a/a_{\rm eq}}}\right) \dot{\delta}_{\rm idm} = S(\eta),\label{eq:idm_eq}
\end{equation}
where $S(\eta) = - 3 \phi''+k^2 \phi - \frac{3}{\eta} \phi'$ is the term sourcing the growth, $\phi$ is the Newtonian potential (we have ignored anisotropic stress), and we have assumed that in our regime of interest $\delta_{\rm idm} \gg \delta_{\rm DR}$. During radiation domination, the Hubble friction is increased by $1+\kappa$. Once we have $z<{\rm min}(z_t,z_{\rm eq})$ the interaction rate drops so that modes that enter the horizon after ${\rm min}(z_t,z_{\rm eq})$ will evolve as in a $\Lambda$CDM universe: $\delta_{\rm idm} \propto a^1$. This implies that the suppression of the matter power spectrum will be on comoving wavenumbers $k>k_c \simeq H_0\sqrt{\Omega_R} \cdot {\rm min}(z_t,z_{\rm eq})$, where $\Omega_R$ is the total radiation energy density in units of the critical energy density. 

It is straightforward to show that during radiation domination and on scales where $\delta \rho_R \gtrsim \delta \rho_M$ in $\Lambda$CDM we have 
\begin{equation}
    \delta_{\rm cdm,\Lambda{\rm CDM}} \sim \log(k\eta)~, \label{app:eq:delta_cdm_rad}
\end{equation}
whereas in the weak model, this gets modified to 
\begin{align}
     \delta_{\rm idm,{\rm weak}} &= \frac{1}{\kappa} +\frac{3^{\kappa/2}(3-2\kappa)(k\eta)^{-\kappa}\cos(\kappa \pi/2) \Gamma(\kappa-1)}{3-\kappa}\nonumber \\
     &\sim \log(k\eta)-\frac{1}{2}\kappa \log^2(k\eta)~,\label{app:eq:weak}
\end{align} 
where we have solved \cref{eq:idm_eq} in the $k\eta \gg 1$ limit using the potential sourced by radiation perturbations, 
\begin{equation}
    \phi = 3 \frac{\sin(k\eta/\sqrt{3})-k\eta/\sqrt{3}\cos(k \eta/\sqrt{3})}{(k\eta/\sqrt{3})^3}~.
\end{equation}
In the second line of \cref{app:eq:weak} we take the $\kappa \ll 1$ limit. 
As we approach ${\rm min}(z_t,z_{\rm eq})$ the strength of the interaction drops, and the modes all start to grow as standard CDM. 

On the other hand, the main impact of the strongly interacting model is the response of the non-interacting cold dark matter to the presence of the strongly coupled fraction, $f_{\rm idm}$\,. While the interacting dark matter is strongly interacting with the dark radiation, it does not cluster. The non-interacting cold dark matter follows the standard equation of motion 
\begin{equation}
    \ddot{\delta}_{\rm cdm,{\rm strong}}+\mathcal{H} \dot{\delta}_{\rm cdm,{\rm strong}} = S(\eta)~.
\end{equation}
While the perturbations are dominated by radiation, the source term, $S$, is the same as it is in $\Lambda$CDM, leading to a standard evolution. After matter domination and on scales where the matter perturbations dominate we have
\begin{equation}
    k^2 \phi \simeq - \frac{3}{2} H_0^2 \frac{(1-f_{\rm idm})\Omega_M}{a} \delta_{\rm cdm,{\rm strong}}~.
\end{equation}
In this case, in $\Lambda$CDM we have 
\begin{equation}
    \delta_{\rm cdm,\Lambda{\rm CDM}} \sim a^1~,\label{eq:delta_cdm_mat}
\end{equation}
whereas for the strongly interacting case 
\begin{equation}
    \delta_{\rm cdm,{\rm strong}} \sim a^{+\frac{1}{4}(\sqrt{25-24f_{\rm idm}}-1)}~, \label{app:eq:strong}
\end{equation}
which then transitions to $\delta_{\rm cdm} \sim a^1$ for $z<z_{\rm idm,dec}$ as at this point the strong interaction rate drops to zero.

\subsection{Impact on matter power spectrum} \label{app:impact_pm}

A comparison between \cref{app:eq:weak,app:eq:strong} shows us that the weak model will produce a suppression that increases logarithmically with $k$ (see also Ref.~\cite{Buen-Abad:2017gxg}), whereas modes that enter the horizon before or around matter/radiation equality will be suppressed by the same amount in the strong model, leading to a relatively constant suppression for modes that enter at $z\gg z_{\rm idm, dec}$. 

The scale at which the matter power spectrum suppression begins for the weak model is directly related to ${\rm min}(z_t,z_{\rm eq})$: $k_c \simeq 1/{\rm max}(\tau_t,\tau_{\rm eq})$. On the other hand, in the strong case, there is a significant delay between $z_t$ and the redshift at which the interacting dark matter decouples from the dark radiation, $z_{\rm idm, dec}$. We can estimate the difference between these two redshifts by noting that the interaction rate in the strong model decreases mainly due to the exponential term [see~\cref{eq:strong_Gamma}], giving 
\begin{equation}
    \frac{z_t}{z_{\rm idm,dec}} \simeq \ln\frac{\Gamma^{\rm strong}}{H}\bigg|_{a\ll a_t}~,
\end{equation}
where we have used the fact that $T \propto 1+z$. Given the value of $\alpha_d = 10^{-4}$ and interacting dark matter mass, $m_\chi = 10^3\ {\rm GeV}$, we have $\ln[\Gamma^{\rm strong}/H\big|_{a\ll a_t}] \simeq 20$ leading to $z_{\rm idm,dec} \simeq z_t/20$. The tight coupling between the interacting dark matter and dark radiation leads to a slight delay between when a mode enters the horizon and when the suppression begins. Empirically we find that the suppression occurs after approximately the first half oscillation of the dark radiation, so at scales smaller than $k_c \simeq \pi/\tau_{\rm idm,dec}$.

We are now able to provide approximate scaling laws for the level of suppression in the two models. 
The amplitude of the suppression of the matter power spectrum in the weak model will scale as
\begin{align}
    \frac{P_m^{\rm weak}}{P_m^{\Lambda{\rm CDM}+{\rm DR}}} &= \left(\frac{\delta_{\rm idm}}{\delta_{\rm cdm,\Lambda{\rm CDM}}}\right)^2 \label{app:eq:Pm_weak}\\
    &\sim \begin{cases}
      1-\kappa \ln(k \eta_t)~, &k \eta_t\geq 1~, \\
      1~, &k \eta_t< 1~.
\end{cases}\nonumber
\end{align}
where we have used \cref{app:eq:delta_cdm_rad,app:eq:weak}. The suppression in the strong case is more complicated. First, while the interacting dark matter is tightly coupled, all CDM modes will grow at the suppressed rate given by \cref{app:eq:strong}. Once the interacting dark matter decouples, its density contrast will grow and the dynamics are driven by the combined cold and interacting dark matter system which are coupled through the gravitational potential. Although we can get a qualitative sense of the subsequent evolution (see for example Appendix C of Ref.~\cite{Bernal:2020vbb}), a quantitative approximation is complicated (e.g., Ref.~\cite{Hu:1994uz,lesgourgues:2013}). Here we supply a fitting formula for the asymptotic suppression in the matter power spectrum which captures both the scaling with $f_{\rm idm}$ and $z_t$:
\begin{align}
    \frac{P_m^{\rm strong}}{P_m^{\Lambda{\rm CDM}+{\rm DR}}}\bigg|_{k\gg k_c} &= \left(\frac{\delta_{\rm cdm,{\rm strong}}}{\delta_{\rm cdm,\Lambda{\rm CDM}}}\right)^2\label{app:eq:Pm_strong}\\
    &\approx \left[1+0.43 \cdot f_{\rm idm} \ln\left(\frac{z_t}{3.5\cdot 10^6}\right)\right]^2~, \nonumber
\end{align}

\pagebreak[40]
In the strongly interacting model, the tightly coupled fraction will also impart dark acoustic oscillations of wavelength $\lambda_\mathrm{DAO} \sim 2\pi / [\eta_\mathrm{idm,dec}/\sqrt{3}]$ with an amplitude that scales as $\delta_{\rm DR}/\delta_{\rm cdm,{\rm strong}} \propto z_{\rm idm,dec}/z_{\rm eq}$ (see for example Appendix~C of Ref.~\cite{Bernal:2020vbb}). 

We show the suppression of the matter power spectrum (solid curves) for several choices of $\log_{10}(z_t)$ along with the approximate scalings in \cref{fig:comp_Pm} (dotted curves). There we can see that the logarithmic suppression in the weak model, \cref{app:eq:Pm_weak}, is a very good approximation (left panel). As shown by the dotted vertical lines in the right panel, the scale above which the strong model is suppressed is well-approximated by $k_c \simeq \pi/\tau_{\rm idm,dec}$. The horizontal dotted lines show that the level of the suppression in the strong model is in good agreement with \cref{app:eq:Pm_strong}, but the blue curves ($\log_{10}(z_{\rm idm,dec}) = 3.6$) show this degrades as $z_{\rm idm,dec}$ approaches $z_{\rm rec}$\,. Finally, we can see that for the strong model, the amplitude of the DAO increases as $z_t$ increases. 

\subsection{Impact on the CMB}\label{app:impact_cmb}
\enlargethispage*{2\baselineskip}
\begin{figure}
    \centering
    \includegraphics[width=\columnwidth]{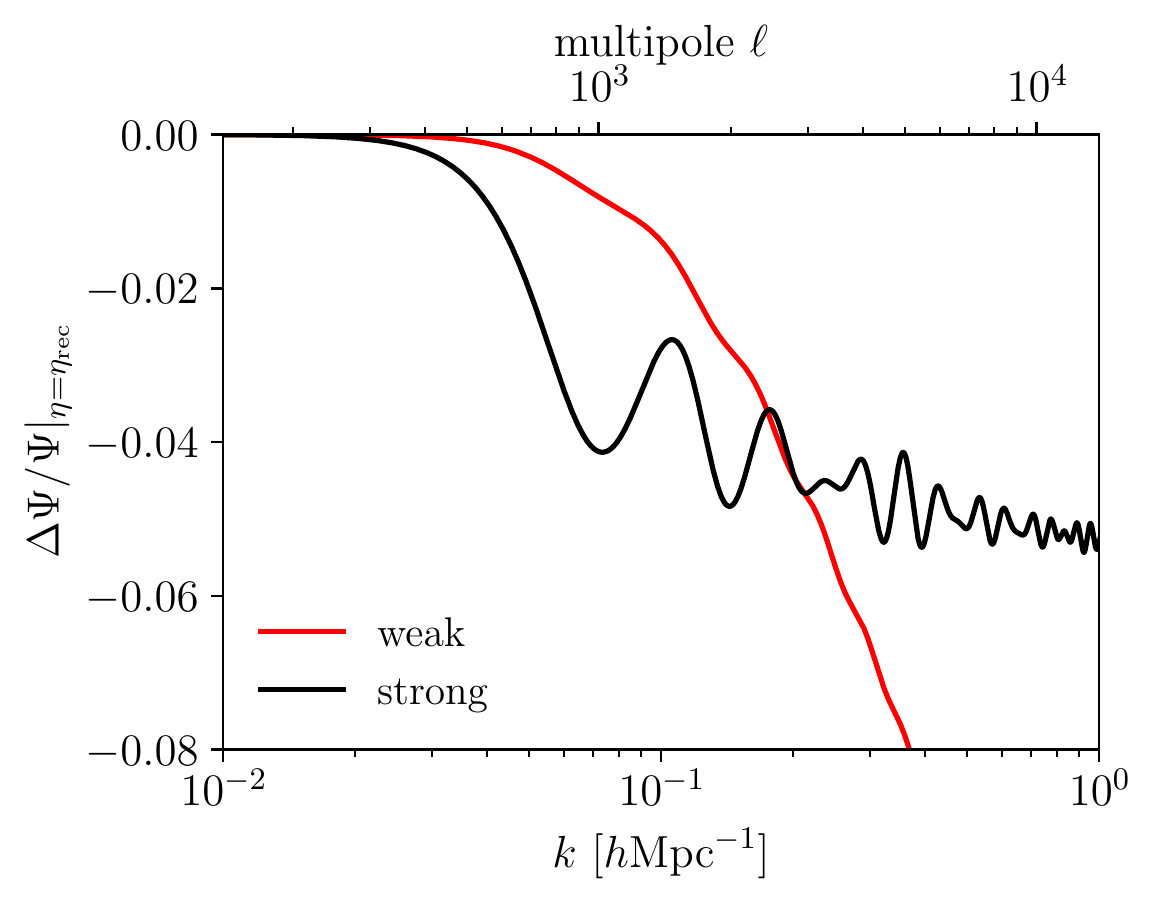}
    \caption{The fractional change, relative to the corresponding model without dark matter-dark radiation interactions, in the Weyl potential evaluated at photon decoupling. The parameters are the same as in \cref{fig:Pm}.}
    \label{fig:weyl}
\end{figure}

Both the weak and strong models imprint a significant suppression to the gravitational potentials by photon decoupling, as shown in \cref{fig:weyl}. In that figure, we have related the wavenumber, $k$, to its corresponding multipole, $\ell$, at the surface of last scattering through $\ell \simeq k (\eta_0-\eta_{\rm dec})$, where $\eta_0$ and $\eta_{{\rm dec}}$ are the conformal time today and at photon decoupling, respectively. Since the dynamics of the interacting dark matter is imprinted on the CMB through gravitational effects, it would appear as though both models will produce CMB power spectra which deviate from $\Lambda$CDM by roughly the same amount at $\ell \sim 2000$. However, as we now discuss, the \emph{time evolution} of the gravitational potentials plays a central role and allows us to distinguish between the impact of the weak and strong models. 

\begin{figure}[!t]
    \centering
    \includegraphics[width=1\columnwidth]{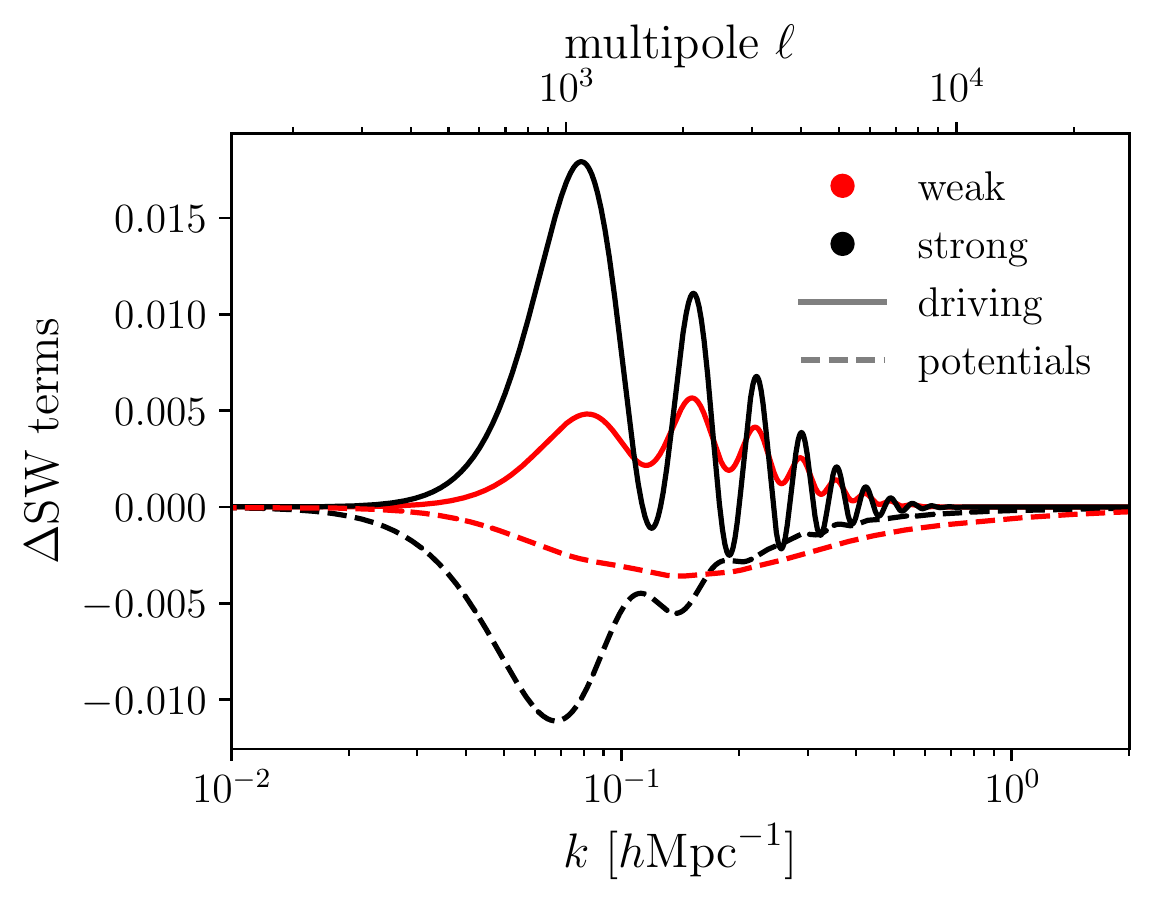}
    \caption{The difference between the driving and potential terms for the weak and strong models for $\log_{10}(z_t) = 4.5$, $N_\mathrm{DR} = 0.5$, $\Gamma_0 = 5.8\times 10^{-7}\ {\rm Mpc}^{-1}$ for the weak model, and $f_{\rm idm} = 0.03$ for the strong model. On the top $x$-axis we have indicated the approximate multipole, $\ell$, through $\ell \simeq k\eta_0$, where $\eta_0$ is the conformal time today.}
    \label{fig:SW-terms}
\end{figure}
Under the tight coupling approximation, the photon fluid equations can be written as a damped, driven, harmonic oscillator with the Weyl potential $\Psi \equiv (\phi+\psi)/2$, playing the role of a driving force. From this, we can write the Sachs-Wolfe contribution to the temperature anisotropies as
\begin{align}
    &\left(\frac{\Delta T (\vec k,\eta)}{T_{\rm CMB}}\right)_{\rm SW} \simeq \zeta(\vec k)\Bigg[e^{-k^2/k_D^2} \Bigg\{ \underbrace{-\cos \left(\frac{k\eta}{\sqrt{3}} \right)}_\text{\clap{free oscillations~}} \nonumber \\ 
    &\underbrace{- \frac{2k}{\sqrt{3}} \int_0^\eta d\eta' \Psi(k,\eta') \sin \left(\frac{k[\eta-\eta']}{\sqrt{3}}\right)}_\text{\clap{driving~}} +\underbrace{\phi(k,\eta)}_\text{\clap{potential}}\Bigg\} \nonumber
    \\ 
    & + \underbrace{\psi(k,\eta)}_\text{\clap{gravitational redshift~}}\Bigg], \label{eq:TTlos}
\end{align}
where $\zeta(\vec k)$ is the primordial curvature perturbation, the exponential factor takes into account the damping effects of photon diffusion which occurs over a length scale $\sim 1/k_D$, and we have approximated the sound horizon as $r_s(\eta) \simeq \eta/\sqrt{3}$. From this, we can see that the Sachs-Wolfe contribution to the temperature anisotropies consists of four terms: a term giving us free oscillations (note that for adiabatic initial conditions, we have used the fact that $\frac{1}{4} \delta_\gamma(k,0) + \phi(k,0) = -1$ \cite{Ma:1995ey}), an integrated term due to the driving effects of the Weyl potential, a term due to the potential $\phi$, and finally the gravitational redshift due to $\psi$. We will use an instantaneous decoupling approximation so the above equation will be evaluated at $\eta = \eta_{\rm rec}$\,.

In order to fit the observed CMB power spectrum we expect deviations from $\Lambda$CDM will be small, in which case the difference in the predicted temperature power spectrum is well approximated by
\begin{align}
    &\Delta C_\ell^{\rm TT} \simeq (4\pi)^2\int k^2d k P_{\rm prim}(k)\ {\rm SW}_{\Lambda{\rm CDM}}(k,\eta_{\rm dec})\times\nonumber \\ &\bigg\{ -\frac{2k}{\sqrt{3}}e^{-k^2/k_D^2}\int_0^{\eta_{\rm dec}} d\eta' \Delta \Psi(k,\eta') \sin \left(\frac{k[\eta_{\rm dec}-\eta']}{\sqrt{3}}\right) \nonumber\\
    & \Delta \phi(k,\eta_{\rm dec}) e^{-k^2/k_D^2} + \Delta \psi(k,\eta_{\rm dec})\bigg\}j^2_\ell(k[\eta_0-\eta_{\rm dec}]),\nonumber \\ \label{eq:approx_diff}
\end{align}
where we have assumed instantaneous photon decoupling at the conformal time $\eta_{\rm dec}$, ${\rm SW}_{\Lambda{\rm CDM}}(k,\eta_{\rm dec})$ is the standard $\Lambda$CDM Sachs-Wolfe terms. We can therefore see that the change in the temperature power spectrum will be a balance between the driving term and the contribution from the gravitational potentials.  

\begin{figure*}[!t]
    \centering
    \includegraphics[width=1\columnwidth]{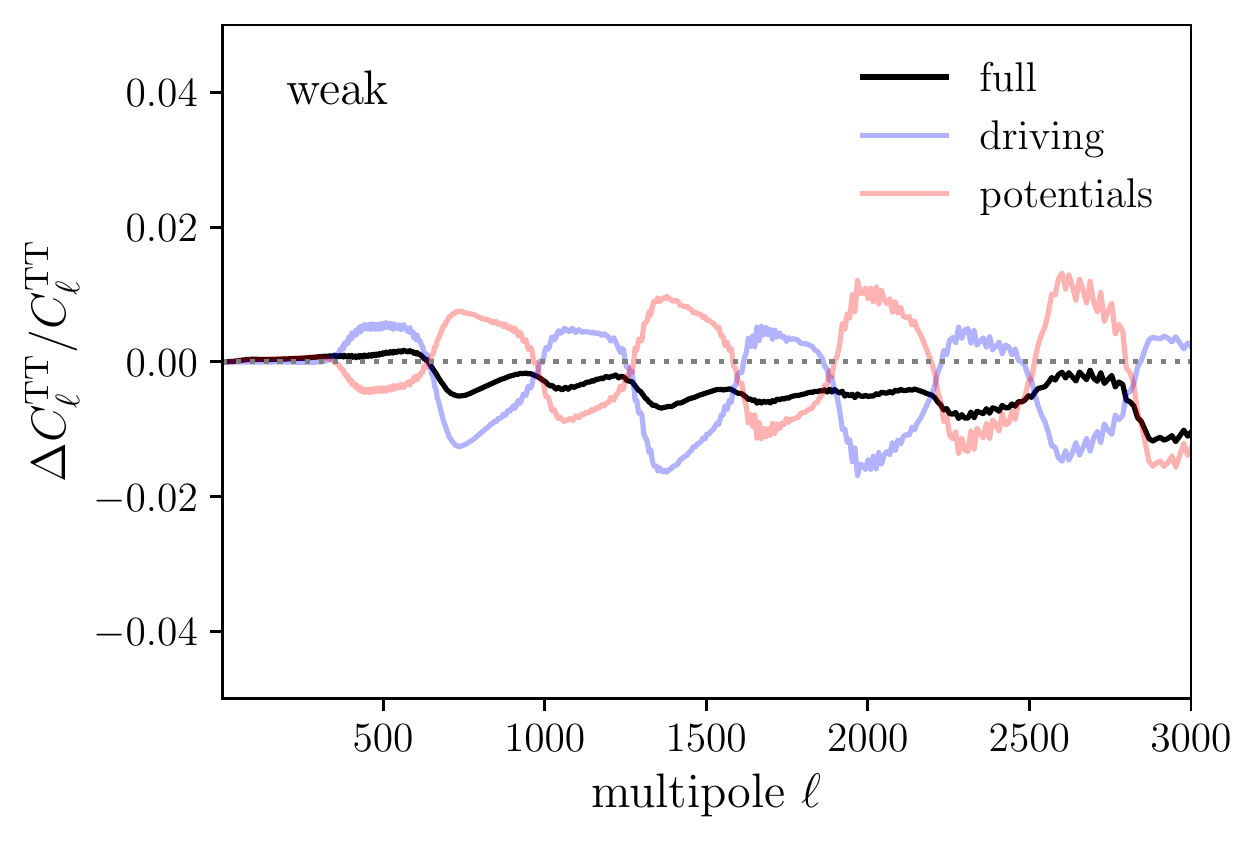}
    \includegraphics[width=0.85\columnwidth]{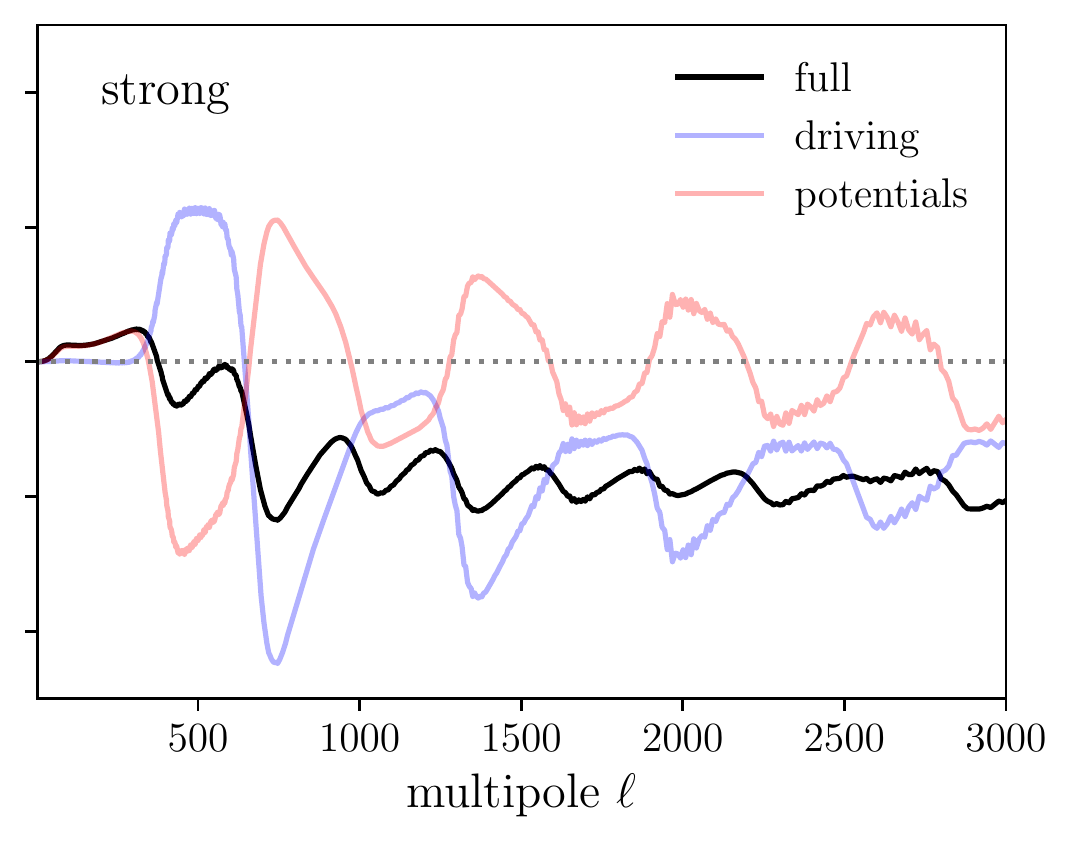}
    \caption{The approximate fractional difference [\cref{eq:approx_diff}] between the temperature power spectrum (only SW effect) for both the weak (\emph{left}) and strong (\emph{right}) models using the same model parameters as in \cref{fig:comp_Pm}. The full difference (black) is the sum of the driving (blue) and potential (red) contributions. A comparison between the differences here and  \cref{fig:Pm}  shows that \cref{eq:approx_diff} is a good approximation to the full calculation.}
    \label{fig:approx-diff}
\end{figure*}

Since for the same value of $z_t$ the strong model maintains interactions for longer compared to the weak model (for example with $\log_{10}(z_t) = 4.5$ we have $\log_{10}(z_{\rm idm,dec}) = 3.1$) we can already anticipate that these contributions will be significantly different between the two models. We confirm this in \cref{fig:SW-terms} where we can see that the strong model produces larger changes to the Sachs-Wolfe terms than in the weak model. It is also important to note that the two contributions have opposite signs so that they will, to some extent, cancel when combined to compute the full $\Delta C_\ell^{\rm TT}$. 

The amount of cancellation between the driving and potential terms determines the overall difference between the predicted temperature power spectra. \cref{fig:approx-diff} shows how these two terms contribute to the overall fractional difference in the temperature power spectrum. For both models, we have set \mbox{$\log_{10}(z_t)=4.5$}. The left panel shows that for the weak model, these terms nearly cancel, leading to a $\lesssim 1\%$-level difference. The right panel shows that for the strong model, the driving term is offset leading to a $\lesssim 2\%$-level difference. Note that \textit{Planck} measures multipoles around $\ell \sim 1000$ to within the cosmic variance limit which corresponds to $\Delta C_\ell/C_\ell = \sqrt{2/(2\ell+1)} \simeq 0.03 (1000/\ell)^{1/2}$, indicating that we should expect \textit{Planck} to be sensitive to such an offset. A comparison between this \cref{fig:approx-diff} (approximate) and \cref{fig:Pm} (full) shows that \cref{eq:approx_diff} is a good approximation to the full calculation.

Even though we have focused on the CMB temperature anisotropies in this Section, in the tight baryon-photon coupled limit the (scalar) CMB polarized anisotropies are computed from the photon dipole moment, which is simply related to the temperature anisotropies through the photon continuity equation. We also note that the early integrated Sachs-Wolfe effect may also differentiate the two models, but will only contribute to scales ($\ell \lesssim \eta_0/\eta_{\rm dec} \sim 100$) larger than those which constrain these models large angular scales.

As discussed in the main text, for values of $f_{\rm idm}$ which may resolve the $S_8$ tension, the delay between $z_t$ and $z_{\rm idm,dec}$ in the strong model leads to a poor fit to the measured CMB power spectra. We can now see this is mainly due to the fact that the strong interaction induces a time evolution in the gravitational potential which, in turn, imprints itself in the CMB through the gravitational driving term. This implies that if we make $z_t$ large enough in the strong model we may be able to limit this time evolution around photon decoupling and provide a better fit to CMB data. Indeed, as we show in \cref{sec:alt_priors}, when we allow $z_t$ to be as large as $10^6$ the strong model is able to resolve both tensions.

\section{Impact of nonlinear corrections}\label{app:nonlinear}

While most of our results depend entirely on predictions that can be obtained starting from linear perturbation theory, for the lensing of the primary CMB anisotropies it is well known that the non-linear enhancement of structure formation has a small but non-zero impact, even in the $\Lambda$CDM model. As such, we use {\sf Halofit} \cite{Smith:2002dz} (taking into account the \cite{Takahashi:2012em} corrections) to estimate the non-linear matter clustering  solely for the purpose of the CMB lensing (we explicitly do not use it for EFTofBOSS/EFTofeBOSS). While this code has been shown to provide excellent estimates of the non-linear clustering in $\Lambda$CDM cosmologies (and some extensions thereof), this estimate could potentially be biased in the interacting stepped dark radiation cosmologies used within this work. Given our summary of minimal $\chi^2$ in \cref{tab:chi2}, we can conclude that the difference for the strongly interacting model due to {\sf Halofit} is minute ($\Delta \chi^2 < 3.5$). For the weakly interacting model, the conclusion is mostly the same ($\Delta \chi^2 < 4.5$), except for the $\mathcal{DS}$ case, where the difference is $6.3$. More importantly, we note that none of the constraints shift by more than $\sim 20\%$ when not including {\sf Halofit}, and this remains true despite the shifts in $\chi^2$, even for the $\mathcal{DS}$ case for the weakly interacting model with the largest $\chi^2$ difference. 

\section{$\chi^2$ and best fit values}\label{app:chi2}

We provide for convenience and reproducibility a table of the minimized $\chi^2$ in \cref{tab:chi2} at the best-fitting points indicated in \cref{tab:bestfit_baseline} for the baseline analysis, minimized using the same algorithm as in \cite{Schoneberg:2021qvd}. We note that these $\chi^2$ are accurate to about $0.5$ in magnitude.

\begin{table*}
    \centering
    \begin{tabular}{c|c c c c| c c c c}
         & \multicolumn{4}{c | }{strongly interacting model}& \multicolumn{4}{c}{weakly interacting model}\\
        \rule{0pt}{4ex} Parameter & $\mathcal{D}$  & $\mathcal{DH}$ & $\mathcal{DS}$& $\mathcal{DHS}$& $\mathcal{D}$  & $\mathcal{DH}$ & $\mathcal{DS}$ & $\mathcal{DHS}$\\ \hline
       \rule{0pt}{3ex} $H_0$[km/s/Mpc]    & 68.49 & 71.50 & 67.94 & 71.90 & 68.09 & 71.70 & 68.24 & 71.80\\
        $S_8$ & 0.825 & 0.832 & 0.784 & 0.798 & 0.794 & 0.824 & 0.776 & 0.781 \\ \hline
        \rule{0pt}{3ex}$10^9A_s$ & 3.053 & 3.049 & 3.046 & 3.031 & 3.038 & 3.051 & 3.038 & 3.053\\
        $n_s$ & 0.9721 & 0.9829 & 0.9716 & 0.9715 & 0.9708 & 0.9820& 0.9733 & 0.9892\\
        $\Omega_m$ & 0.3103 & 0.2993 & 0.3089 & 0.2911 & 0.3067 & 0.2979 & 0.3055 & 0.2984 \\
        $\tau_\mathrm{reio}$ & 0.0573 & 0.0548 & 0.0555 & 0.0554 & 0.0520 & 0.0570 & 0.0519 & 0.0565 \\ \hline
        \rule{0pt}{3ex} $N_\mathrm{DR}$ & 0.13 & 0.64 & 0.01 & 0.74 & 0.01 & 0.64 & 0.02 & 0.64 \\
        $\Gamma_0$ [$10^{-6}$/Mpc] & -- & -- & -- & -- & 1.03 & 0.07 & 1.17 & 0.63 \\
        $f_\mathrm{idm}$ & 0.01 & $4 \times 10^{-5}$ & 0.08 & $4 \times 10^{-4}$ & -- & -- & -- & --\\
    \end{tabular}
    \caption{Best-fit values for the different parameters for the weak and the strong model, given the baseline data set $\mathcal{D}$ and either the $H_0$ prior ($\mathcal{H}$) or $S_8$ prior ($\mathcal{S}$).}
    \label{tab:bestfit_baseline}
\end{table*}

\begin{table*}[t]
    \centering

    \begin{tabular}{c c|c |c c c c c c c c c c}
        Data & Model & $\chi^2$ tot & P18TTTEE & P18lens & BAO &Pantheon & H & S & ACT+SPT & BOSS & eBOSS\\ \hline \rule{0pt}{3ex}
        $\mathcal{D}$ & $\Lambda$CDM & 3806.02 & 2765.83 & 8.80 & 5.50 &1025.87 & -- & -- & -- & -- & --\\
        $\mathcal{D}$ & weak  & 3804.27 & 2762.64 & 10.65 & 5.23 &1025.75 & -- & -- & -- & -- & --\\
        $\mathcal{D}$ & strong & 3805.53 & 2764.85 & 9.27 & 5.54 &1025.88 & -- & -- & -- & -- & --\\
        $\mathcal{D}$ (wo HF) & weak & 3807.51 & 2766.18 & 9.04 & 5.24 & 1027.05 & -- & -- & -- & -- & --\\
        $\mathcal{D}$ (wo HF) & strong & 3806.79 & 2765.36 & 8.85 & 6.42 &1026.15 & -- & -- & -- & -- & --\\ \hline \rule{0pt}{3ex}
        $\mathcal{DH}$ & $\Lambda$CDM & 3838.48 & 2771.39 & 9.33 & 5.62 & 1025.63 & 26.51 & -- & -- & -- & --\\
        $\mathcal{DH}$ & weak & 3812.54 & 2768.23 & 10.07 & 5.81 & 1025.64 & 2.79 & -- & -- & -- & --\\
        $\mathcal{DH}$ & strong & 3814.92 & 2770.17 & 10.81 & 5.51 & 1025.63 & 2.81 & -- & -- & -- & --\\
        $\mathcal{DH}$ (wo HF) & weak & 3814.56  & 2769.62 & 9.71 & 6.05 & 1026.87 & 2.31 & -- & -- & -- & --\\
        $\mathcal{DH}$ (wo HF) & strong & 3815.39 & 2769.53 & 10.14 & 6.50 & 1025.66 & 3.57 & -- & -- & -- & --\\ \hline \rule{0pt}{3ex}
        $\mathcal{DS} $ & $\Lambda$CDM & 3814.35 & 2767.92 & 9.81 & 5.19 & 1025.66 & -- & 5.76 & -- & -- & --\\
        $\mathcal{DS}$ & weak & 3805.00 & 2763.21 & 10.67 & 5.17 & 1025.71 & -- & 0.23 & -- & -- & --\\
        $\mathcal{DS}$ & strong & 3809.04 & 2767.57 & 9.34 & 5.39 & 1025.83 & -- & 0.91 & -- & -- & --\\
        $\mathcal{DS}$ (wo HF) & weak & 3811.29 & 2770.23 & 9.49 & 5.38 & 1025.82 & -- & 0.37 & -- & -- & --\\
        $\mathcal{DS}$ (wo HF) & strong & 3811.72 & 2769.14 & 9.29 & 6.70 & 1026.22 & -- & 0.37 & -- & -- & --\\ \hline \rule{0pt}{3ex}
        $\mathcal{DHS}$ & $\Lambda$CDM & 3841.38 & 2772.32 & 10.30 & 6.43 & 1025.66 & 23.73 & 2.94 & -- & -- & --\\ 
        $\mathcal{DHS}$ & weak & 3814.90 & 2770.34 & 10.17 & 5.72 & 1025.63 & 2.41 & 0.63 & -- & -- & --\\ 
        $\mathcal{DHS}$ & strong & 3821.55 & 2771.76 & 10.79 & 7.85 & 1025.83 & 2.98 & 2.38 & -- & -- & --\\
        $\mathcal{DHS}$ (wo HF) & weak & 3819.46 & 2772.74 & 10.44 & 5.76 & 1026.86 & 2.07 & 1.59 & -- & -- & --\\
        $\mathcal{DHS}$ (wo HF) & strong & 3824.93 & 2773.15 & 11.55 & 7.06 & 1025.70 & 3.35 & 4.22 & -- & -- & --\\ \hline \rule{0pt}{3ex}
        $\mathcal{DHS}$ (high $z_t$) & strong & 3817.74 & 2773.73 & 9.84 & 5.60 & 1025.63 & 2.85 & 0.10 & -- & -- & --\\ 
        $\mathcal{DHS}$ (free $f_{\rm idm}$) & weak & 3815.68 & 2770.42 & 9.90 & 5.9 & 1025.63 & 2.82 & 1.01 & -- & -- & --\\
        $\mathcal{DHS}$ (free $r_g$) & strong & 3815.95 & 2769.11 & 10.41 & 5.49 & 1025.63 & 0.67 & 4.64 & -- & -- & --\\ \hline \rule{0pt}{3ex}
        $\mathcal{DHS}$+ACT+SPT & $\Lambda$CDM & 5220.18 & 2778.96 & 10.30 & 6.00 & 1026.05 & 24.93 & 3.40 & 1370.54 & -- & --\\ 
        $\mathcal{DHS}$+ACT+SPT & weak & 5186.57 & 2772.57 & 9.79 &5.44 & 1025.63& 6.45 & 0.56 & 1366.13 & -- & --\\
        $\mathcal{DHS}$+ACT+SPT & strong & 5195.75  & 2769.75 & 10.57  & 8.12  &  1027.00 &  4.52 & 6.43  &  1369.36 & -- & -- \\
        \hline \rule{0pt}{3ex}
        $\mathcal{D}$+EFT & $\Lambda$CDM & 4022.69 &  2765.80 & 8.87 & 1.41$^{(*)}$ & 1025.78 & -- & -- & -- & 159.69 & 61.13\\
        $\mathcal{D}$+EFT & weak &  4021.53 & 2764.73 & 9.04 & 1.49$^{(*)}$ & 1025.74 & -- & -- & -- & 159.50 & 61.02\\
        $\mathcal{D}$+EFT & strong & 4020.90 & 2766.56 & 8.82 & 1.36$^{(*)}$ & 1025.81 & -- & -- & -- & 157.90 & 60.46\\ \hline \rule{0pt}{3ex}
        $\mathcal{DH}$+EFT & $\Lambda$CDM & 4055.67 & 2774.01 & 9.80 & 2.44$^{(*)}$ & 1025.72 & 24.64 & -- & -- & 158.33 & 60.71 \\
        $\mathcal{DH}$+EFT & weak & 4029.56 & 2767.57 & 9.75 & 2.51$^{(*)}$ & 1025.64 & 3.58 & -- & -- & 159.69 & 60.80\\
        $\mathcal{DH}$+EFT & strong & 4038.03 & 2772.17 & 9.43 & 2.34$^{(*)}$ & 1025.63 & 4.55 & -- & -- & 158.75 & 61.58\\ \hline \rule{0pt}{3ex}
        $\mathcal{DS}$+EFT & $\Lambda$CDM & 4035.11 & 2774.69 & 9.84 & 2.22$^{(*)}$ &  1025.62 & -- &  3.40  & -- & 158.17 & 61.17\\
        $\mathcal{DS}$+EFT & weak & 4023.97 & 2765.78 & 9.24 & 1.64$^{(*)}$ &  1025.69 & -- &  1.08  & -- & 159.75 & 60.79\\ 
        $\mathcal{DS}$+EFT & strong & 4033.13 & 2769.75 & 10.09 &1.62$^{(*)}$ & 1025.70 & -- &  2.24  & -- & 158.72 & 61.49\\ \hline \rule{0pt}{3ex}
        $\mathcal{DHS}$+EFT & $\Lambda$CDM & 4059.04 &  2774.04 & 9.79 & 2.43$^{(*)}$ & 1025.73 & 24.66 & 3.33 & -- & 158.34 & 60.72\\ 
        $\mathcal{DHS}$+EFT & weak & 4033.78 &  2769.14 & 9.80 & 2.46$^{(*)}$ & 1025.64 & 3.84 & 1.99 & -- &  159.93  & 60.99\\ 
        $\mathcal{DHS}$+EFT & strong & 4044.31 & 2773.86 & 10.98 & 3.10$^{(*)}$ & 1025.78 & 2.26 &  3.24  & -- & 159.14 & 61.22\\ \hline \rule{0pt}{3ex}
        $\mathcal{DHS}$+ACT+SPT+EFT & weak & 5401.92 & 2771.23 & 9.88 & 2.3$^{(*)}$ & 1025.63 & 5.64 &  1.09  & 1365.87 & 159.36 & 60.87\\ \hline \rule{0pt}{3ex}
    \end{tabular}
    \caption{Table of the best-fit $\chi^2$ for various combinations of model and datasets, separated by individual likelihood contributions. $(*)$ We note that the `BAO' for analyses including EFTofBOSS/EFTofeBOSS (denoted as BOSS/eBOSS in the header) only refer to the low-z data from 6dFGS and SDSS DR7, but does not include BOSS DR12.
    }
    \label{tab:chi2}
\end{table*}

\end{appendix}
\end{document}